\documentclass[onecolumn]{aa}
\usepackage{amsmath,amssymb,amscd}
\usepackage{graphicx}
\usepackage{txfonts}
\begin{document}
\title{Tidal dynamics of extended bodies\\
in planetary systems and multiple stars}

\author{S. Mathis\inst{1,2}, C. Le Poncin-Lafitte\inst{3}}

\offprints{S. Mathis}

\institute
{Laboratoire AIM, CEA/DSM - CNRS - Universit\'e Paris Diderot, IRFU/Service d'Astrophysique, CEA-Saclay, F-91191 Gif-sur-Yvette Cedex, France\\
\email{stephane.mathis@cea.fr} \and
LUTH, Observatoire de Paris - CNRS - Universit\'e Paris-Diderot; Place Jules Janssen, F-92195 Meudon Cedex, France \and
SYRTE UMR8630, Observatoire de Paris, 61 Avenue de l'Observatoire, F-75014 Paris, France\\
\email{christophe.leponcin-lafitte@obspm.fr}\\
}

\date{Received ; accepted }

\abstract
{With the discovery during the past decade of a large number of extrasolar planets orbiting their parent stars at distances lower than 0.1 astronomical unit (and the launch and the preparation of dedicated space missions such as CoRoT -Bord\'e {\it et al.}, 2003- and KEPLER -Borucki {\it et al.}, 2007-), with the position of inner natural satellites around giant planets in our Solar System and with the existence of very close but separated binary stars, tidal interaction has to be carefully studied.} 
{This interaction is usually studied with a ponctual approximation for the tidal perturber. The purpose of this paper is to examine the step beyond this traditional approach by considering the tidal perturber as an extended body. To achieve this, the gravitational interaction between two extended bodies and more precisely the interaction between mass multipole moments of their gravitational fields and the associated tidal phenomena are studied.}
{Use of Cartesian Symmetric Trace Free tensors, of their relation with spherical harmonics and of the Kaula's transform enables us to analytically derive the tidal and mutual interaction potentials as well as the associated disturbing functions in extended bodies systems.}
{The tidal and mutual interaction potentials of two extended bodies are derived. In addition, the external gravitational potential of such tidally disturbed extended body is obtained, using the Love number theory, as well as the associated disturbing function. Finally, the dynamical evolution equations for such systems are given in their more general form without any linearization. We also give, under simplified assumption, a comparison of this formalism to the ponctual case. {\bf The non-ponctual terms have to be taken into account for strongly deformed perturbers ($J_{2}\ge10^{-2}$) in very close systems ($a_{\rm B}/R_{\rm B}\le5$).}}
{We show how to derive explicitely the dynamical equations for the gravitational and tidal interactions between extended bodies and associated dynamics. Conditions for application of such formalism are given.}

\maketitle

\section{Introduction}
\par
\begin{figure}[h!]
\centering
\resizebox{12cm}{!}{\includegraphics{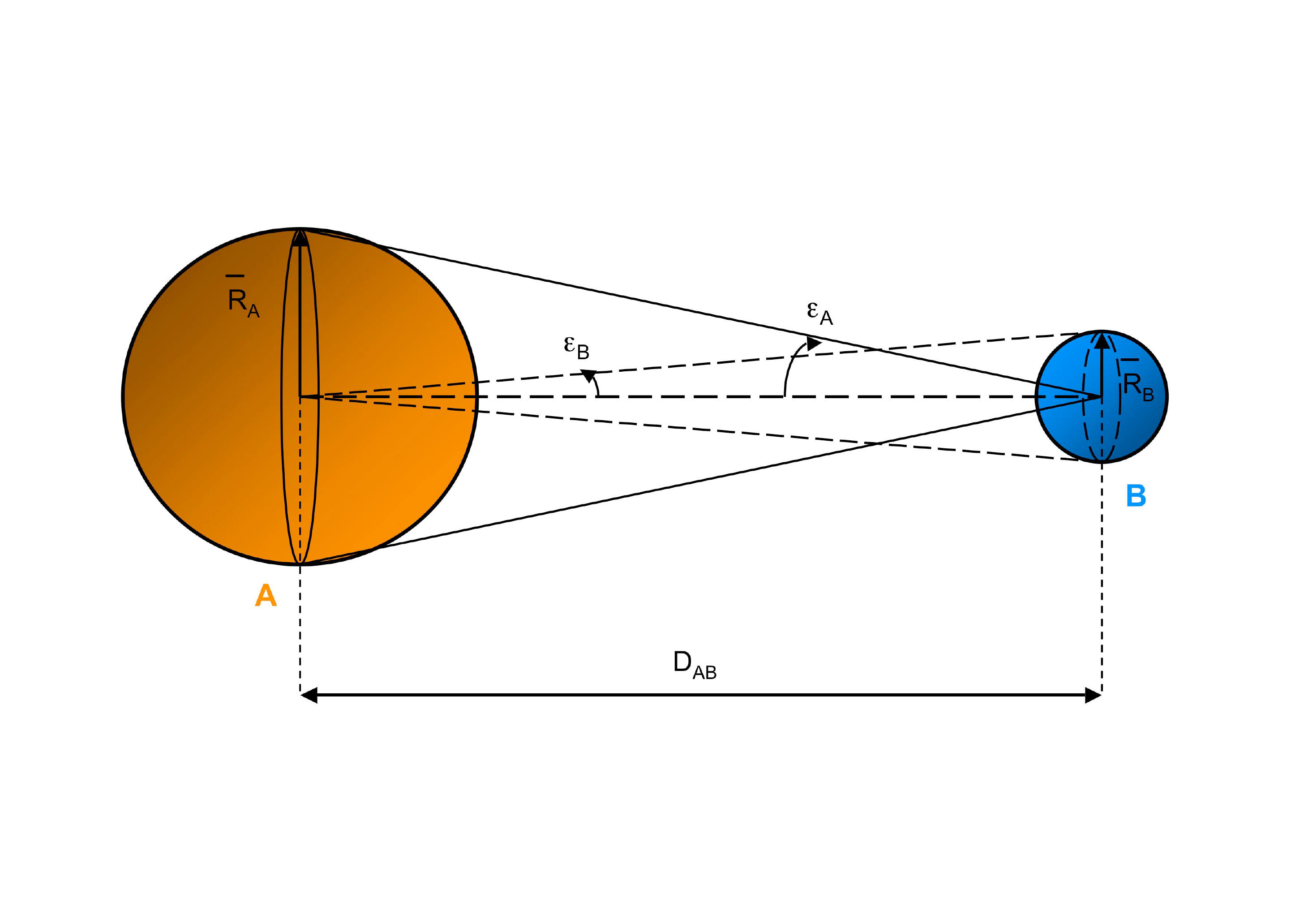}}
\caption{System of two extended bodies. ${\overline R}_{\rm A}$ and ${\overline R}_{\rm B}$ are the respective mean radius of A and B while $D_{\rm AB}$ is the distance between their respective center of mass. $\varepsilon_{\rm A}$ and $\varepsilon_{\rm B}$ are the conical angles with which each perturber is seen from the center of mass of the perturbed body. When $\frac{{\overline R}_{i}}{D_{\rm AB}}\!<\!\!<\!1$ with $i=$A or B the i$^{\rm th}$ body could be considered as a ponctual mass perturber.}
\label{Perturber}
\end{figure}

 In celestial mechanics,  one of the main approximation done in the modelling of tidal effects (star-star, star-innermost planet or planet-natural satellites interactions) is to consider the tidal perturber as a point mass body. However a large number of extrasolar Jupiter-like planets orbiting their parent stars at a distance lower than 0.1 AU have been discovered during the past decade (Mayor {\it et al.}, 2005). Moreover, in Solar System, Phobos around Mars and the inner natural satellites of Jupiter, Saturn, Uranus and Neptune are very close to their parent planets. In such cases, the ratio of the perturber mean radius to the distance between the center of mass of the bodies can be not any more negligeable compared to 1 (cf. Fig.~\ref{Perturber}). Furthermore, it can be also the case for very close but separated binary stars. In that situation, neglecting the extended character of the perturber have to be relaxed, so the tidal interaction between two extended bodies must be solved in a self-consistent way with taking into account the full gravitational potential of the extended perturber, generally expressed with some mass multipole moments, and then to consider their interaction with the tidally perturbed body. In the litterature, not so many studies have been done (Borderies, 1978-1980; Ilk, 1983; Borderies \& Yoder, 1990; Hartmann, Soffel \& Kioustelidis, 1994; Maciejewski, 1995). The purpose of this work is then to provide a theoretical procedure to obtain this tidal gravitational interaction as well as its associated tidal dynamical evolution.\\

\par Several years ago, Hartmann, Soffel \& Kioustelidis (1994) introduced in Celestial Mechanics an interesting tool, based on Cartesian Symmetric Trace Free (STF) tensors, to treat straighforwardly the couplings between the gravitational fields of extended bodies. These tensors are fully equivalent to usual spherical harmonics but in addition a set of STF tensors represents an irreductible basis of the rotation group SO3 (Courant \& Hilbert 1953, Gelfand {\it et al.} 1963). It means that using algebraic properties of STF tensors with the index notation of Blanchet \& Damour (1986), these objects become a powerful tool to determine the coupling between spherical harmonics in an elegant and compact way. However, as these tensors are not widely used in Celestial mechanics, we first recall their definition and fundamental properties and we stress their relation with usual spherical harmonics. Then, we treat the multipole expansion of gravitational-type fields for which each type is related to a given extended body. First, the well-known external field of such body is derived using STF tensors; classical identities are provided. Next, the mutual gravitational interaction between two extended bodies and the associated tidal interaction are derived. We show how the use of STF tensors leads to an analytical and compact treatment of the coupling of their gravitational fields. We deduce the general expressions of tidal and mutual interaction potentials expanded in spherical harmonics. Using the classical Kaula's transform (Kaula, 1962), we express them as a function of the Keplerian orbital elements of the body considered as the tidal perturber. These results are used to derive the external gravitational potential of such tidally perturbed extended body. Introducing a third body, its mutual interaction potential with the previous tidally perturbed extended body is defined that allows us to derive the disturbing function using the results obtained with STF tensors and the Kaula's transform. At this stage, the different type of mutual gravitational interaction are defined. The dynamical equations ruling the evolution of this system are obtained. Finally, we use a reduced form of that equations to qualitatively quantify the influence of non-ponctual terms of the disturbing function in comparison with the ponctual case.

\section{STF-Multipole expansion of gravitational potentials}

\subsection{Definitions and notations}

\par We focus first on the STF tensors. Let us define a Cartesian l-tensor as a set of numbers $T_{i_1i_2...i_l}$ with $l$ different indices $i_1$ to $i_l$, each taking integer value running between 1 and 3. A compact multi-index notation, first introduced by Blanchet \& Damour (1986), is generally used. An uppercase latin letter denotes a multi-index while the corresponding lowercase denotes its number of indices :
\begin{equation}
L=i_1i_2...i_l\, , \quad T_L=T_{i_1i_2...i_l}\, .
\end{equation}
The Einstein summation convention is assumed in the following, so if some index appears twice, a summation over that index is implied
\begin{equation}
A_KB_K=A_{i_1i_2...i_k}B_{i_1i_2...i_k}=\sum_{i_1,i_2,...,i_k}A_{i_1i_2...i_k}B_{i_1i_2...i_k}\, .
\end{equation}
Given a Cartesian tensor ${\cal A}_L$, we denote its symmetric part with parenthesis
\begin{equation}
\label{SYMformula}
{\cal A}_{(L)}={\cal A}_{(i_1...i_l)}=\frac{1}{l!}\sum_\sigma {\cal A}_{i_{\sigma(1)}i_{\sigma(2)}...i_{\sigma(l)}}\, ,
\end{equation}
where $\sigma$ runs over all the $l!$ permutations of $\left\{1,2,3,...,l\right\}$.\\

Next, the Symmetric Trace Free part of a tensor ${\cal A}_L$ is denoted indifferently by $\hat{\cal A}_L={\cal A}_{<L>}={\cal A}_{<i_1i_2...i_l>}$. Following Thorne (1980), the STF part of ${\cal A}_L$ reads
\begin{equation}
\label{STFformula}
\hat{\cal A}_L=\sum_{k=0}^{\left\lbrack\frac{l}{2}\right\rbrack}a^{k,l}\delta_{(i_1i_2}...\delta_{i_{2k-1}i_{2k}}S_{i_{2k+1}...i_l)a_1a_1..a_ka_k}
\end{equation}
where $\delta$ is the classical Kronecker delta function,
\begin{equation}
S_L={\cal A}_{(L)}\quad\hbox{and}\quad
a^{k,l}=\frac{l!}{(2l-1)!!}\frac{(-1)^k}{(2k)!!}\frac{(2l-2k-1)!!}{(l-2k)!}\,,
\end{equation}
$\left\lbrack\frac{l}{2}\right\rbrack$ denoting the integer part of $\frac{l}{2}$ while
\begin{equation*}
l!=l\left(l-1\right)\left(l-2\right)\cdot\cdot\cdot2\times1\quad\hbox{and}\quad l!!=l\left(l-2\right)\left(l-4\right)\cdot\cdot\cdot\times
\left\{
\begin{array}{l@{\quad}l}
1\quad\hbox{if}\,\, l \,\,\hbox{is odd}\\
2\quad\hbox{if}\,\, l \,\,\hbox{is even}
\end{array}
\right.\, .
\end{equation*}

\subsection{STF-basis}

\par Let ${\bf e}_i$ ($i$ running between 1 and 3) be a Cartesian basis vectors set (with the rule ${\bf e}_{i}\cdot{\bf e}_{j}=\delta_{ij}$). The basis of the $(2l+1)$-dimensional vector space of STF rank $l$-tensors is made of the STF parts of the $l$-fold tensorial products (Thorne, 1980)
\begin{equation}
\left\lbrack\bigotimes_{n=1}^m{\bf E}^+\right\rbrack\left\lbrack\bigotimes_{p=m+1}^l{\bf E}^0\right\rbrack
\end{equation}
where
\begin{equation}
{\bf E}^+\equiv {\vec e}_1+i{\vec e}_2\, , \quad {\bf E}^0\equiv {\vec e}_3
\end{equation}
with $i^2=-1$. For $m>0$, let us define the following algebraic object
\begin{equation}
E_L^{lm}=\left\lbrack\prod_{n=1}^{m}E^+_{i_n}\right\rbrack \left\lbrack \prod_{p=m+1}^{l}E^0_{i_p} \right\rbrack\,.
\end{equation}
Then, the STF canonical basis is proportional to $E^{lm}_{<L>}$ and can be chosen as (Thorne, 1980):
\begin{equation}
\label{base1}
\hat{\cal Y}_L^{l,m}=A^{lm}E^{lm}_{<L>},
\end{equation}
where
\begin{equation}
\label{base2}
A^{lm}=(-1)^m(2l-1)!!\sqrt{\frac{2l+1}{4\pi(l-m)!(l+m)!}}\, .
\end{equation}
The constant $A^{lm}$ is choosen to get a normalization such that
\begin{equation}
\label{orthogonality}
\hat{\cal Y}_L^{l,m}\left(\hat{\cal Y}_L^{l,m'}\right)^*=\frac{(2l+1)!!}{4\pi l!}\delta_{mm'}\, ,
\end{equation}
$z^{*}$ corresponds to the complex conjugate of $z$, $z$ being a complex number or function. Finally, taking into account Eqs. (\ref{STFformula}), (\ref{base1}) and (\ref{base2}), we obtain
\begin{equation}
\label{STFbasis-calcul}
\hat{{\cal Y}}_L^{l,m}={\mathcal N}_{lm}\sum_{j=0}^{\left\lbrack\frac{l-m}{2}\right\rbrack}a_{lmj}\delta_{(i_1i_2}...\delta_{i_{2j-1}i_{2j}}E^+_{i_{2j+1}}...E^+_{i_{2j+m}}E^0_{i_{2j+m+1}}...E^0_{i_{l})}\, ,
\end{equation}
where
\begin{equation}
a^{lmj}=\frac{(-1)^j}{2^lj!(l-j)!}\frac{(2l-2j)!}{(l-m-2j)!}
\quad\hbox{and}\quad
{\mathcal N}_{lm}=(-1)^m\sqrt{\frac{2l+1}{4\pi}\frac{(l-m)!}{(l+m)!}}\, .
\end{equation}
Let us consider an arbitrary vector $\vec{x}=(x^1,x^2,x^3)$. We can now introduce the Euclidean norm $r$ of $\vec{x}$, the corresponding unit vector $\vec{n}$ and the Cartesian tensors $x^K$ and $n^K$ constructed on $\vec{x}$  as follow
\begin{equation}
\label{definition}
r=\sqrt{\vec{x}^2}\, ,\quad \vec{n}=\vec{x}/r\, ,\quad x^K=x^{i_1}x^{i_2}....x^{i_k}\, , \quad n^K=n^{i_1}n^{i_2}....n^{i_k}\, .
\end{equation}
Using the harmonic property $\nabla^{2}\left(r^{-1}\right)\equiv 0$ (for $r>0$) and Eq. (\ref{definition}), one gets
\begin{equation}
\label{derlaplace}
\hat{\partial}_L\left(\frac{1}{r}\right)=\partial_L\left(\frac{1}{r}\right)=(-1)^l(2l-1)!!\frac{\hat{n}_L}{r^{l+1}},
\end{equation}
where $\hat{\partial}_L$ and $\hat{n}_L$ are the STF part of $\partial_L$ and $n^L$, respectively. Then, considering Eq. (\ref{STFbasis-calcul}) and taking into account $\nabla^{2}\left(r^{-1}\right)\equiv 0$, the computation of Kronecker deltas functions combined with Eq. (\ref{derlaplace}) leads to
\begin{eqnarray}
\hat{\cal Y}_L^{l,m}\partial_L\left(\frac{1}{r}\right)&=&{\mathcal N}_{lm}a^{lm0}(\partial_x+i\partial_y)^m\partial_z^{l-m}\left(\frac{1}{r}\right)\nonumber\\
&=&(-1)^l(2l-1)!!\frac{\hat{\cal Y}_L^{l,m}\hat{n}_L}{r^{l+1}}\, .
\label{Id1}
\end{eqnarray}
By applying twice Eq. (\ref{Id1}), we find
\begin{eqnarray}
\hat{\cal Y}^{l,m}_L\hat{\partial}_L\left[\hat{\cal Y}^{j,k}_J\hat{\partial}_J\left(\frac{1}{r}\right)\right]&=&\hat{\cal Y}^{l,m}_L\hat{\partial}_L\left\lbrack {\mathcal N}_{jk}a^{jk0}(\partial_x+i\partial_y)^k\partial_z^{j-k}\left(\frac{1}{r}\right)\right\rbrack\nonumber\\
&=&{\mathcal N}_{jk}a^{jk0}(\partial_x+i\partial_y)^k\partial_z^{j-k}\left\lbrack\hat{\cal Y}^{l,m}_L\hat{\partial}_L\left(\frac{1}{r}\right)\right\rbrack\, .
\end{eqnarray}
This last equation leads to a composition law of a product of basis functions $\hat{\cal Y}^{l,m}_L$. After some algebra, we obtain
\begin{equation}
\label{composition1}
\hat{\cal Y}^{l,m}_L\hat{\partial}_L\left[\hat{\cal Y}^{j,k}_J\hat{\partial}_J\left(\frac{1}{r}\right)\right]=(-1)^{l+j}(2l-1)!!(2j-1)!!\frac{\gamma_{j,k}^{l,m}}{r^{l+j+1}}\hat{\cal Y}_{LJ}^{l+j,m+k}\hat{n}_{LJ}
\end{equation}
where
\begin{equation}
\gamma_{jk}^{lm}=\gamma_{lm}^{jk}=\sqrt{\frac{2l+1}{(l+m)!(l-m)!}\frac{2j+1}{(j+k)!(j-k)!}\frac{\left[\left(l+j\right)-\left(m+k\right)\right]!\left[\left(l+j\right)+\left(m+k\right)\right]!}{4\pi\left[2\left(l+j\right)+1\right]}}\, .
\end{equation}
Taking into account that the left hand side of the Eq. (\ref{composition1}) can be written using Eq. (\ref{derlaplace}) as follow 
\begin{equation}
\hat{\cal Y}^{l,m}_L\hat{\partial}_L\left[\hat{\cal Y}^{j,k}_J\hat{\partial}_J\left(\frac{1}{r}\right)\right]=\left(-1\right)^{l+j}\left(2l+2j-1\right)!!\hat{\cal Y}_L^{l,m}\hat{\cal Y}_J^{j,k}\frac{\hat{n}_{LJ}}{r^{l+j+1}}\, ,
\end{equation}
we finally get
\begin{equation}
\hat{\cal Y}_{L}^{l,m}\hat{\cal Y}_{J}^{j,k}\hat{n}_{LJ}=\frac{(2l-1)!!(2j-1)!!}{(2l+2j-1)!!}\gamma_{jk}^{lm}\hat{\cal Y}_{LJ}^{l+j,m+k}\hat{n}_{LJ}\, .\label{couplageYLm}
\end{equation}

\subsection{Relation between STF-basis and usual spherical harmonics}

In this section the equivalence between STF basis and spherical harmonics $Y_{l,m}$ is given. Following Abramowitz and Stegun (1970), $Y_{l,m}$ reads
\begin{eqnarray}
\label{harmo1}
Y_{l,m}(\theta,\varphi)&=&{\mathcal N}_{lm}P_{l}^{m}(\cos\theta)\exp\left[im\varphi\right]\quad\hbox{for $m\ge 0$}\nonumber\\
&=&{\mathcal N_{lm}}\left(\exp\left[i\varphi\right]\sin\theta\right)^m\sum_{j=0}^{\left\lbrack\frac{l-m}{2}\right\rbrack}a^{lmj}(\cos\theta)^{l-m-2j}\, ,
\end{eqnarray}
where the $P_{l}^{m}$ are the classical associated Legendre Polynomials. We also recall the symmetry property of the $Y_{l,m}$, namely:
\begin{equation}
Y_{l,-m}\left(\theta,\varphi\right)=\left(-1\right)^{m}Y_{l,m}^{*}\left(\theta,\varphi\right)\quad\hbox{with $m\ge0$}.
\label{Ysym}
\end{equation}

Since the components of the unit vector $\vec{n}$ in complex form as follow
\begin{equation}
\label{harmo2}
n_x+in_y=\exp\left[i\varphi\right]\sin\theta\, , \quad n_z=\cos\theta\, ,
\end{equation}
we can identify
\begin{equation}
\label{Ylm2STF}
Y_{l,m}=\hat{\cal Y}^{l,m}_Ln_L\equiv\hat{\cal Y}^{l,m}_L\hat{n}_L\, ,
\end{equation} 
which can be inverted by using Eq. (\ref{orthogonality}) as
\begin{equation}
\label{nlmYlm}
\hat{n}_L=\frac{4\pi l!}{(2l+1)!!}\sum_{m=-l}^{+l}\hat{\cal Y}_L^{l,m}Y_{l,m}^{*}\, .
\end{equation}
To give an example of how it works, let us examine the cases $l=0$ and $l=1$. The spherical harmonics are
\begin{equation}
Y_{0,0}=\frac{1}{\sqrt{4\pi}}\, ,
\end{equation}
\begin{equation}
Y_{1,0}=\sqrt{\frac{3}{4\pi}}\cos\theta\, , \quad Y_{1,1}=-\sqrt{\frac{3}{8\pi}}\sin\theta \exp\left[i\varphi\right]\, .
\end{equation}
Let us compare with the STF-basis functions. For $l=0$, the normalization rule given in Eq. (\ref{orthogonality}) gives the single number $1/\sqrt{4\pi}$. For $l=1$ we have
\begin{equation}
\left( \begin{array}{c}
\hat{\cal Y}_1^{1,0}  \\
\hat{\cal Y}_2^{1,0} \\
\hat{\cal Y}_3^{1,0}  \end{array} \right)=\sqrt{\frac{3}{4\pi}}
\left( \begin{array}{c}
0  \\
0  \\
1\end{array} \right)\quad\hbox{while}\quad
\left( \begin{array}{c}
\hat{\cal Y}_1^{1,1}  \\
\hat{\cal Y}_2^{1,1} \\
\hat{\cal Y}_3^{1,1}  \end{array} \right)=-\sqrt{\frac{3}{8\pi}}
\left( \begin{array}{c}
1  \\
i  \\
0\end{array} \right)\quad\hbox{and}\quad
\left( \begin{array}{c}
\hat{\cal Y}_1^{1,-1}  \\
\hat{\cal Y}_2^{1,-1} \\
\hat{\cal Y}_3^{1,-1}  \end{array} \right)=\sqrt{\frac{3}{8\pi}}
\left( \begin{array}{c}
1  \\
-i \\
0\end{array} \right)\, ;
\end{equation}
it is verified that $Y_{1m}=\hat{\cal Y}_{i}^{1m}(x^i/r)$.

\subsection{Multipole expansion of the external gravitational field of an extended body}

\begin{figure}[h!]
\centering
\resizebox{12cm}{!}{\includegraphics{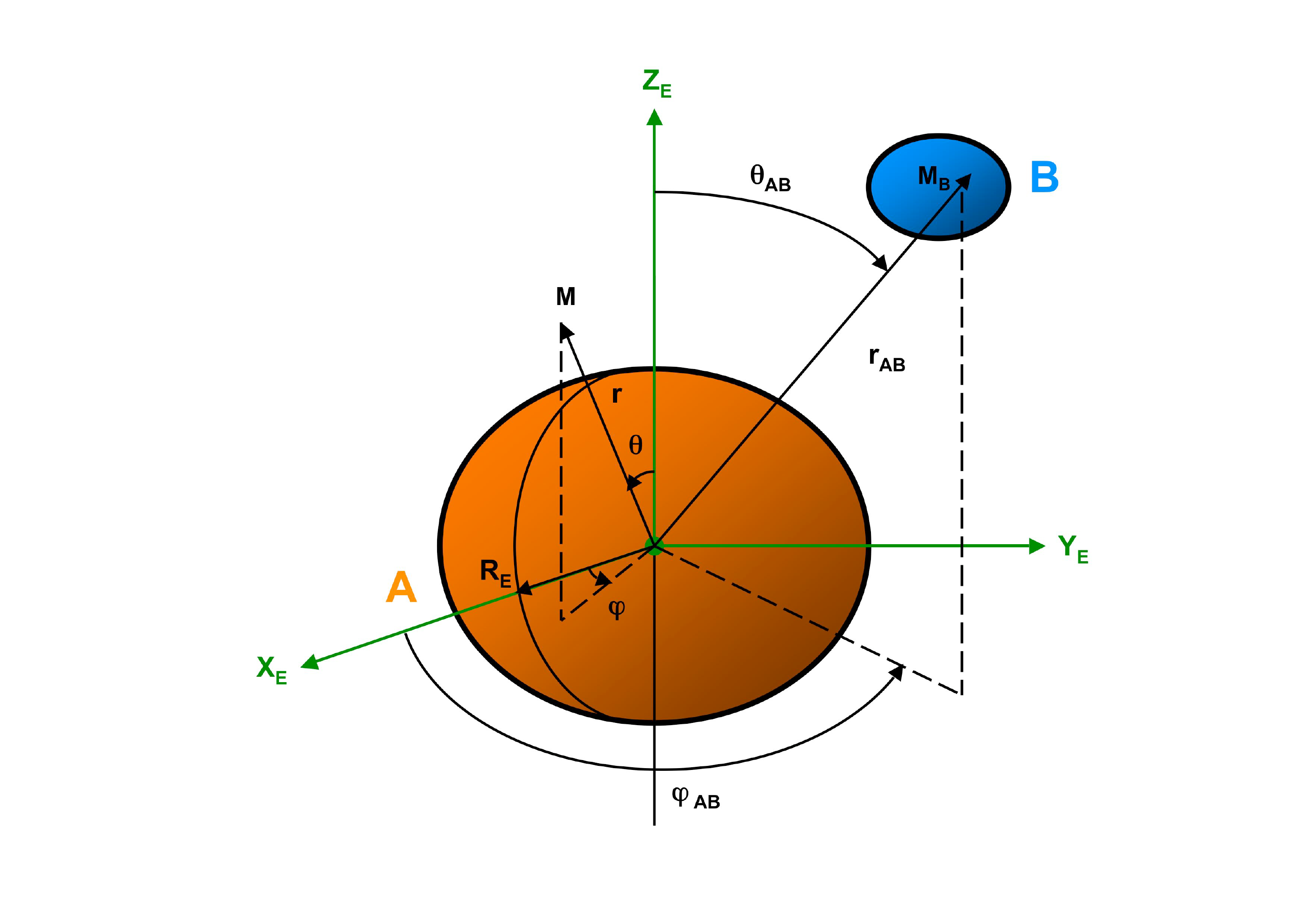}}
\caption{Spherical coordinates system associated to the equatorial reference frame ${\mathcal R}_{\rm E}:\left\{O_{\rm A}, {\rm X}_{\rm E}, {\rm Y}_{\rm E}, {\rm Z}_{\rm E}\right\}$ of an extended body A; we have $\vec r\equiv\left(r,\theta,\varphi\right)$ and $\vec r_{\rm AB}\equiv\left(r_{\rm AB},\theta_{\rm AB},\varphi_{\rm AB}\right)$ where $r_{\rm AB}$, $\theta_{\rm AB}$ and $\varphi_{\rm AB}$ are the coordinates of the center of mass of the potential extended perturber B. $R_{\rm A}$ is the equatorial radius of A.}
\label{StructureTide}
\end{figure}

Let us consider some matter distribution, corresponding to a body $A$, in the inertial coordinates $(t,x^i)$. The Newtonian gravitational potential of this body, $V^{\rm A}\left(t,{\vec x}\right)$, is obtained by solving the Poisson equation
\begin{equation}
\nabla^{2} V^{\rm A}\left(t,{\vec x}\right)=-4\pi G \rho_{\rm A}\left(t,{\vec x}\right)\quad\hbox{with}\quad\lim_{\vert{\vec x}\vert\rightarrow\infty} V^{\rm A}(t,{\vec x})=0\, ,
\end{equation}
$\rho_{\rm A}\left(t,\vec x\right)$ being its density. This leads an expression of $\vert\vec x\vert\ge R_{\rm A}$, where $R_{\rm A}$ is the equatorial radius of A, to
\begin{equation}
V^{\rm A}(t,{\vec x})=G\int_A \frac{\rho_{\rm A}(t,{\vec x'})}{\vert {\vec x}-{\vec x'}\vert}{\rm d}^3{\vec x'}\, .
\end{equation}
Then, the external gravitational field of the body A for $\vert\vec r\vert\ge R_{\rm A}$ can be represented by the series:
\begin{equation}
V^{\rm A}(t,{\bf x})=G\sum_{l_{\rm A}=0}^{\infty}\frac{(-1)^{l_{\rm A}}}{l_{\rm A}!}\hat{\cal M}_{L_{\rm A}}\hat{\partial}_{L_{\rm A}}\left(\frac{1}{r}\right)\, ,
\label{GravityA}
\end{equation}
where all mass multipole moments are defined by:
\begin{equation}
\hat{\cal M}_{L_{\rm A}}=\frac{l_{\rm A}!}{(2l_{\rm A}-1)!!}\sum_{m_{\rm A}=-l_{\rm A}}^{l_{\rm A}}M_{l_{\rm A},m_{\rm A}}\hat{\cal Y}_{L_{\rm A}}^{l_{\rm A},m_{\rm A}}
\label{MlmSTF}
\end{equation}
the usual gravitational moments in the physical space being given by:
\begin{equation}
M_{l_{\rm A},m_{\rm A}}=\frac{4\pi}{2l_{\rm A}+1}\int_{M_{\rm A}} r^{l_{\rm A}} Y_{l_{\rm A},m_{\rm A}}^{*}\left(\theta,\varphi\right){\rm d}M_{\rm A};
\label{Mdef}
\end{equation}
$M_{\rm A}$ is the mass of A and ${\rm d}M_{\rm A}=\rho_{A}r^2{\rm d}r\sin\theta{\rm d}\theta{\rm d}\varphi$. Inserting Eq. (\ref{MlmSTF}) into Eq. (\ref{GravityA}) and using Eq. (\ref{Id1}), the final expression for $V^{\rm A}$ for $\vert\vec r\vert\ge R_{\rm A}$ is thus obtained:
\begin{equation}
V^{\rm A}\left(t,\vec r\right)=G\sum_{l_{\rm A}=0}^{\infty}\sum_{m_{\rm A}=-l_{\rm A}}^{l_{\rm A}}M_{l_{\rm A},m_{\rm A}}\frac{Y_{l_{\rm A},m_{\rm A}}\left(\theta,\varphi\right)}{r^{l_{\rm A}+1}}.
\label{UGA}
\end{equation}
One should note the symmetry property of $M_{l_{\rm A},m_{\rm A}}$:
\begin{equation}
M_{l_{\rm A},-m_{\rm A}}=\left(-1\right)^{m_{\rm A}}M_{l_{\rm A},m_{\rm A}}^{*}.
\label{sym}
\end{equation}
Moreover, $M_{l_{\rm A},m_{\rm A}}$ could be represented in its polar form
\begin{equation}
M_{l_{\rm A},m_{\rm A}}=|M_{l_{\rm A},m_{\rm A}}|\exp\left[i\delta M_{l_{\rm A},m_{\rm A}}\right],
\end{equation}
where the following identities are obtained from Eq. (\ref{sym})
\begin{equation}
\left\{
\begin{array}{l@{\quad}l}
|M_{l_{\rm A},-m_{\rm A}}|=|M_{l_{\rm A},m_{\rm a}}|\\
{\rm Arg}\left(M_{l_{\rm A},-m_{\rm A}}\right)=m_{\rm A}\pi-{\rm Arg}\left(M_{l_{\rm A},m_{\rm A}}\right)
=m_{\rm A}\pi-\delta M_{l_{\rm A},m_{\rm A}}
\end{array}
\right. .
\end{equation}
Using the classical symmetry property concerning spherical harmonics given in Eq. (\ref{Ysym}), $V^{\rm A}\left(t,\vec r\right)$ could also be expressed with the associated Legendre polynomials:
\begin{equation}
V^{\rm A}\left(t,\vec  r\right)=G\sum_{l_{\rm A}=0}^{\infty}\sum_{m_{\rm A}=0}^{l_{\rm A}}\frac{P_{l_{\rm A}}^{m_{\rm A}}\left(\cos\theta\right)}{r^{l_{\rm A}+1}}\left[C_{l_{\rm A},m_{\rm A}}\cos\left(m_{\rm A}\varphi\right)+S_{l_{\rm A},m_{\rm A}}\sin\left(m_{\rm A}\varphi\right)\right],
\end{equation}
where the usual coefficients $C_{l_{\rm A},m_{\rm A}}$ and $S_{l_{\rm A},m_{\rm A}}$ are given by:
\begin{equation}
\left\{
\begin{array}{l@{\quad}l}
C_{l_{\rm A},m_{\rm A}}={\mathcal N}_{l_{\rm A}}^{m_{\rm A}}\left(2-\delta_{m_{\rm A},0}\right){\rm R}_{\rm e}\left(M_{l_{\rm A},m_{\rm A}}\right)\\
S_{l_{\rm A},m_{\rm A}}=-2{\mathcal N}_{l_{\rm A}}^{m_{\rm A}}\left(1-\delta_{m_{\rm A},0}\right){\rm I}_{\rm m}\left(M_{l_{\rm A},m_{\rm A}}\right)
\end{array}
\right. .
\end{equation}
The expression of $M_{l_{\rm A},m_{\rm A}}$ and $\delta M_{l_{\rm A},m_{\rm A}}$ are then deduced for $m_{\rm A}\ge 0$:
\begin{equation}
|M_{l_{\rm A},m_{\rm A}}|=\frac{1}{{\mathcal N}_{l_{\rm A}}^{m_{\rm A}}}\sqrt{\left[\frac{C_{l_{\rm A},m_{\rm A}}}{\left(2-\delta_{m_{\rm A},0}\right)}\right]^2+\left(\frac{S_{l_{\rm A},m_{\rm A}}}{2}\right)^2\left(1-\delta_{m_{\rm A},0}\right)^2},
\end{equation}
\begin{equation}
\delta M_{l_{\rm A},m_{\rm A}}=-{\rm Arctan}\left[\frac{\left(1-\delta_{m_{\rm A},0}\right)\left(2-\delta_{m_{\rm A},0}\right)}{2}\frac{S_{l_{\rm A},m_{\rm A}}}{C_{l_{\rm A},m_{\rm A}}}\right].
\end{equation}
{In the general case, the gravitational moments are expanded as:
\begin{equation}
M_{l_{\rm A},m_{\rm A}}=M_{l_{\rm A},m_{\rm A}}^{\rm S_{\rm A}}+M_{l_{\rm A},m_{\rm A}}^{\rm T_{\rm A}}\,.
\label{GME}
\end{equation}
$M_{l_{\rm A},m_{\rm A}}^{\rm S_{\rm A}}$ and $M_{l_{\rm A},m_{\rm A}}^{\rm T_{\rm A}}$ are respectively those in the case where A is isolated (without any perturber) and those induced by the tidal perturber(s).}\\

One can identify some special values of $M_{l_{\rm A},m_{\rm A}}$ relevant for the gravitational field of a body A. The trivial one is its mass, $M_{\rm A}$
\begin{equation}
M_{0,0}=\sqrt{4\pi}M_{\rm A}.
\end{equation}
Furthermore, we know that the external field of an axisymmetric body A can be expressed as a function of the usual multipole moment $J_{l_{\rm A}}$\footnote{They are driven by two types of deformation. The first one is those induced by internal dynamical processes such that rotation (through the centrifugal acceleration) and magnetic field (through the volumetric Lorentz force). The second one is the axisymmetric permanent tidal oval shape due to a companion in close binary or multiple systems.} (see {\it e. g.} Roxburgh 2001)
\begin{equation}
V^{\rm A}\left(t,\vec r\right)=\frac{GM_{\rm A}}{r}\left[1-\sum_{l_{\rm A}>0}J_{l_{\rm A}}\left(\frac{R_{\rm A}}{r}\right)^{l_{\rm A}}P_{l_{\rm A}}\left(\cos\theta\right)\right];
\end{equation}
using Eq. (\ref{UGA}), we identify in a straigthforward way:
\begin{equation}
V^{\rm A}\left(t,\vec r\right)=G\left[\frac{M_{\rm A}}{r}+\sum_{l_{\rm A}=0}^{\infty}M_{J_{l_{\rm A}};l_{\rm A},0}^{\rm A}\frac{Y_{l_{\rm A},0}\left(\theta,\varphi\right)}{r^{l_{\rm A}+1}}\right]\quad\hbox{where}\quad M_{J_{l_{\rm A}};l_{\rm A},0}^{\rm A}=M_{l_{\rm A},0}^{\rm S_{\rm A}}+M_{l_{\rm A},0}^{\rm T_{\rm A}}=-\frac{J_{l_{\rm A}} M_{\rm A} R_{\rm A}^{l_{\rm A}}}{{\mathcal N}_{l_{\rm A}}^{0}}.
\end{equation}
We can now focus on the second type of gravitational interaction, namely the tides between two extended bodies.

\subsection{Determination of the tidal potential}
Let us now introduce an accelerated reference frame, {\it i.e.} $(t,X_{\rm A}^i)$, associated with a body A which is related to a global inertial frame through the transformation
\begin{equation}
x^i=z_{\rm A}^i(t)+X_{\rm A}^i\, ,
\end{equation}
$z_{\rm A}^i(t)$ being the arbitrary motion of the local A-frame. The equations of motion with respect to the local A-frame reads (Damour, Soffel, Xu 1993):
\begin{equation}
\label{motion1}\frac{\partial \rho_{\rm A}}{\partial t}+\frac{\partial (\rho_{\rm A} v_{\rm A}^i)}{\partial X_{\rm A}^i}=0\, ,
\end{equation}
 \begin{equation}
\label{motion2}\frac{\partial (\rho_{\rm A}v_{\rm A}^i)}{\partial t}+\frac{\partial}{\partial X_{\rm A}^j} (\rho_{\rm A}
v_{\rm A}^iv_{\rm A}^j+t^{ij})=\rho_{\rm A}\frac{\partial V_{\rm eff}^{\rm A}}{\partial X^i_{\rm A}}\, ,
\end{equation}
where $\rho_{\rm A}(t,{\vec X}_{\rm A})\equiv\rho_{\rm A}(t,{\vec z}_{\rm A})$ is the mass volumic density expressed in the local A-frame, $v_{\rm A}^i$ being the velocity with respect to this frame while $t^{ij}$ denotes the stress tensor. The following effective potential appears
\begin{equation}
\label{evident}
V_{\rm eff}^{\rm A}(t,{\vec X}_A)=\sum_{B=1}^{N}V^{\rm B}(t,{\vec z}_{\rm A}+{\vec X}_{\rm A})-V_{\rm ext}^{\rm A}(t,{\vec z}_{\rm A})-\frac{{\rm d}^2{\vec z}_{\rm A}}{{\rm d}t^2}.{\vec X}_{\rm A}\, ,
\end{equation}
where
\begin{equation}
V_{\rm ext}^{\rm A}(t,\vec{X}_{\rm A})=\sum_{{\rm B}\ne{\rm A}}V^{\rm B}(t,\vec{X}_{\rm A})\, ,
\end{equation}
the considered body A being tidally interacting with $N-1$ perturbing extended bodies B; $V^{\rm B}$ is the potential of each body B different from A.
The last term of Eq. (\ref{evident}) represents the inertial effects on the accelerated local frame A. This effective potential can be split into the potential of A given in Eq. (\ref{UGA}) and a tidal potential, $V_{\rm T}^{\rm A}$, as follow:
\begin{equation}
V_{\rm eff}^{\rm A}=V^{\rm A}+V^{\rm A}_{\rm T}\, ,
\end{equation}
the tidal part being given by
\begin{equation}
V^{\rm A}_{\rm T}(t,{\vec X}_A)=V_{\rm ext}^{\rm A}(t,{\vec z}_{\rm A}+{\vec X}_{\rm A})-V_{\rm ext}^{\rm A}(t,{\vec z}_{\rm A})-\frac{{\rm d}^2{\vec z}_{\rm A}}{{\rm d}t^2}\cdot{\vec X}_{\rm A}\,.
\end{equation} 
By integrating over the body A the equations of motion Eqs. (\ref{motion1}) and (\ref{motion2}), we get the equations for the conservation of the total mass of A, $M^{\rm A}=M_{\rm A}$, and for the second time derivative of the local dipole moment $M_{\rm A}^i$
(Damour {\it et al.} 1992, Hartmann {\it et al.} 1994):
\begin{equation}
\frac{{\rm d}M^{\rm A}}{{\rm d}t}=0\, ,
\end{equation}
\begin{equation}
\frac{{\rm d}^2M_i^{\rm A}}{{\rm d}t^2}=\int_{\rm A}\rho_{\rm A}\frac{\partial V_{\rm T}^{\rm A}}{\partial X_{\rm A}^i}d^3\vec{X}_{\rm A}\, .
\label{TideEquation}
\end{equation}
We can now derive the expression of $V_{\rm T}^{\rm A}$ into series by using an STF expansion of ascending powers of ${\vec X}_{\rm A}$
as follow
\begin{equation}
\label{tideDef1}
V_{\rm T}^{\rm A}=\sum_{B\ne A}\sum_{l_{\rm A}=1}^\infty\frac{1}{l_{\rm A}!}\hat{X}_{\rm A}^{L_{\rm A}}G_{L_{\rm A}}^{\rm A}\, ,
\end{equation}
where $G^{\rm A}_{L_{\rm A}}$ are the local effective tidal moments with
\begin{equation}
\label{tideDef1}G_i^{\rm A}=\partial_i V_{\rm ext}^{\rm A}({\vec z}_{\rm A})-\frac{{\rm d}^2z_{\rm A}^i}{{\rm d}t^2}\, ,
\end{equation}
\begin{equation}
\label{tideDef2}G_{i_1...i_l}^{\rm A}=\partial_{i_1...i_l} V_{\rm ext}^{\rm A}({\vec z}_{\rm A})\, .
\end{equation}
We assume that the origin of the local A-frame coincide with the center of mass of A, {\it i.e.} the dipole moment $M_i^{\rm A}(t)$ vanishes. Since now the $\left\{l,m\right\}$ indices are related to their associated body: for example for A we use $\left\{l_{\rm A},m_{\rm A}\right\}$ while for B we use $\left\{l_{\rm B},m_{\rm B}\right\}$. With these definitions, the right-hand side of Eq. (\ref{TideEquation}) can be written as:
\begin{equation}
\int_{\rm A}\rho_{\rm A}\frac{\partial V_{\rm T}^{\rm A}}{\partial X_{\rm A}^i}d^3\vec{X}_{\rm A}=\sum_{B\ne A}\sum_{l_{\rm A}=0}^\infty\frac{1}{l_{\rm A}!}\hat{\cal
M}^{\rm A}_{L_{\rm A}} G^{\rm A}_{iL_{\rm A}}\, .\label{arnaque}
\end{equation}
The local equation of motion is finally obtained from the d'Alembert Principle with
\begin{equation}
\frac{{\rm d}^2}{{\rm d}t^2}M_i^{\rm A}(t)=0\, ,
\end{equation}
which leads, by using Eqs. (\ref{arnaque}) and (\ref{tideDef1}) , to:
\begin{equation}
M_{\rm A}\frac{{\rm d}^2z_{\rm A}^i}{{\rm d}t^2}=M_{\rm A}\partial_i V_{\rm ext}^{\rm A}+\sum_{B\ne A}\sum_{l_{\rm A}=1}^\infty\frac{1}{l_{\rm A}!}\hat{\cal M}_{L_{\rm A}}^{\rm A}G_{iL_{\rm A}}^{\rm A}\, .
\end{equation}
Expressing $V^B$ in term of Eq. (\ref{GravityA}) as in Hartmann, Soffel \& Kioustelidis (1994), we finally obtain
\begin{equation}
\label{inertial}
M_{\rm A}\frac{{\rm d}^2z_A^i}{{\rm d}t^2}=\sum_{B\ne A}\left\lbrack \sum_{l_{\rm A}=0}^\infty\sum_{l_{\rm B}=0}^\infty\frac{\left(-1\right)^{l_{\rm B}}}{l_{\rm A}!l_{\rm B}!}\hat{\cal M}_{L_{\rm A}}^{\rm A}\hat{\cal M}_{L_{\rm B}}^{\rm B}\partial_{iL_{\rm A}L_{\rm B}}\left(\frac{1}{r_{\rm AB}}\right)\right\rbrack\, .
\end{equation}
Using Eqs. (\ref{tideDef1})-(\ref{tideDef2}) and (\ref{inertial}), the tidal potential can be expressed as:
\begin{equation}
V^{\rm A}_{\rm T}=G\sum_{B\ne A}\left\lbrace \sum_{l_{\rm A}=2}^{\infty}\sum_{l_{\rm B}=0}^{\infty}\frac{(-1)^{l_{\rm B}}}{l_{\rm A}!l_{\rm B}!}\hat{\cal M}^{\rm B}_{L_{\rm B}}\hat{X}_{\rm A}^{L_{\rm A}}\hat{\partial}_{L_{\rm A}L_{\rm B}}\left(\frac{1}{r_{\rm AB}}\right)-\frac{1}{M_{\rm A}}\sum_{l_{\rm A}=1}^\infty\sum_{l_{\rm B}=0}^{\infty}\frac{(-1)^{l_{\rm B}}}{l_{\rm A}!l_{\rm B}!}\hat{\cal M}^{\rm A}_{L_{\rm A}}\hat{\cal M}^{\rm B}_{L_{\rm B}}X_{\rm A}^i\partial_{iL_{\rm A}L_{\rm B}}\left(\frac{1}{r_{\rm AB}}\right)\right\rbrace\,.
\label{general}
\end{equation}
Using Eq. (\ref{derlaplace}), we get
\begin{equation}
V^{\rm A}_{\rm T}=G\sum_{B\ne A}\left\lbrace\sum_{l_{\rm A}=2}^{\infty}\sum_{l_{\rm B}=0}^{\infty}(-1)^{l_{\rm A}}\frac{(2l_{\rm A}+2l_{\rm B}-1)!!}{l_{\rm A}!l_{\rm B}!}\hat{\cal M}^{\rm B}_{L_{\rm B}}\hat{X}_{\rm A}^{L_{\rm A}}\frac{\hat{n}_{L_{\rm A}L_{\rm B}}}{r_{\rm AB}^{l_{\rm A}+l_{\rm B}+1}}-\frac{1}{M_{\rm A}}\sum_{l_{\rm A}=1}^\infty\sum_{l_{\rm B}=0}^{\infty}(-1)^{l_{\rm A}}\frac{(2l_{\rm A}+2l_{\rm B}-1)!!}{l_{\rm A}!l_{\rm B}!}\hat{\cal M}^{\rm A}_{L_{\rm A}}\hat{\cal M}^{\rm B}_{L_{\rm B}}X_{\rm A}^i\partial_i\left(\frac{\hat{n}_{L_{\rm A}L_{\rm B}}}{r_{\rm AB}^{l_{\rm A}+l_{\rm B}+1}}\right)\right\rbrace\, ,
\label{general1}
\end{equation}
that leads, taking into account Eq. (\ref{MlmSTF}), to:
\begin{eqnarray}
V^{\rm A}_{\rm T}&=&G\sum_{B\ne A}\left\lbrace \sum_{l_{\rm A}=2}^{\infty}\sum_{l_{\rm B}=0}^{\infty}\sum_{m_{\rm B}=-l_{\rm B}}^{l_{\rm B}}(-1)^{l_{\rm A}}\frac{(2l_{\rm A}+2l_{\rm B}-1)!!}{(2l_{\rm B}-1)!!l_{\rm A}!}M^{\rm B}_{l_{\rm B},m_{\rm B}}\hat{\cal Y}^{l_{\rm B},m_{\rm B}}_{\rm B}\hat{X}_{\rm A}^{L_{\rm A}}\frac{\hat{n}_{L_{\rm A}L_{\rm B}}}{r_{\rm AB}^{l_{\rm A}+l_{\rm B}+1}}\right.\nonumber\\
&&\left.-\frac{1}{M_{\rm A}}\sum_{l_{\rm A}=1}^\infty\sum_{m_{\rm A}=-l_{\rm A}}^{l_{\rm A}}\sum_{l_{\rm B}=0}^{\infty}\sum_{m_{\rm B}=-l_{\rm B}}^{l_{\rm B}}(-1)^{l_{\rm A}}\frac{(2l_{\rm A}+2l_{\rm B}-1)!!}{(2l_{\rm A}-1)!!(2l_{\rm B}-1)!!}M^{\rm A}_{l_{\rm A},m_{\rm A}}M^{\rm B}_{l_{\rm B},m_{\rm B}}
X_{\rm A}^i\hat{\cal Y}^{l_{\rm A},m_{\rm A}}_{L_{\rm A}}\hat{\cal Y}^{l_{\rm B},m_{\rm B}}_{L_{\rm B}}\partial_i\left(\frac{\hat{n}_{L_{\rm A}L_{\rm B}}}{r_{\rm AB}^{l_{\rm A}+l_{\rm B}+1}}\right)\right\rbrace\, .
\label{general2}
\end{eqnarray}
We now replace the Cartesian multipole moments by their spherical harmonics representation. First of all, using Eq. (\ref{nlmYlm}) and putting $\vec{X}_{\rm A}\equiv \vec r \equiv(r,\theta,\varphi)$, we get:
\begin{equation}
X_{\rm A}^i=\frac{4r \pi}{3}\sum_{m_{\rm A}=-1}^{1}\hat{\cal Y}_i^{1,m_{\rm A}}Y_{l_{\rm A},m_{\rm A}}^{*}(\theta,\varphi),
\end{equation}
\begin{equation}
\hat{X}_{\rm A}^{L_{\rm A}}=\frac{4r^{l_{\rm A}}\pi l_{\rm A}!}{(2l_{\rm A}+1)!!}\sum_{m_{\rm A}=-l_{\rm A}}^{l_{\rm A}}\hat{\cal Y}_{L_{\rm A}}^{l_{\rm A},m_{\rm A}}Y_{l_{\rm A},m_{\rm A}}^{*}(\theta,\varphi)\, .
\end{equation}
Then, the last term of Eq. (\ref{general2}) involving the product of basis function $\hat{\cal Y}$ can be written by using Eq. (\ref{couplageYLm}) as follow:
\begin{eqnarray}
\hat{\cal Y}^{l_{\rm A},m_{\rm A}}_{L_{\rm A}}\hat{\cal Y}^{l_{\rm B},m_{\rm B}}_{L_{\rm B}}\partial_i\left(\frac{\hat{n}_{L_{\rm A}L_{\rm B}}}{r_{\rm AB}^{l_{\rm A}+l_{\rm B}+1}}\right)&=&\partial_i\left(\hat{\cal Y}^{l_{\rm A},m_{\rm A}}_{L_{\rm A}}\hat{\cal Y}^{l_{\rm B},m_{\rm B}}_{L_{\rm B}}\frac{\hat{n}_{L_{\rm A}L_{\rm B}}}{r_{\rm AB}^{l_{\rm A}+l_{\rm B}+1}}\right)\nonumber\\
&=&\frac{(2l_{\rm A}-1)!!(2l_{\rm B}-1)!!}{(2l_{\rm A}+2l_{\rm B}-1)!!}\gamma_{l_{\rm B},m_{\rm B}}^{l_{\rm A},m_{\rm A}}\hat{\cal Y}_{L_{\rm A}L_{\rm B}}^{l_{\rm A}+l_{\rm B},m_{\rm A}+m_{\rm B}}\partial_i\left(\frac{\hat{n}_{L_{\rm A}L_{\rm B}}}{r_{\rm AB}^{l_{\rm A}+l_{\rm B}+1}}\right)\, .\nonumber\\\label{derivdefou}
\end{eqnarray}
Noting that the STF derivative in the RHS of Eq. (\ref{derivdefou}) can be split into two parts
\begin{equation}
\partial_i\left(\frac{\hat{n}_{L_{\rm A}L_{\rm B}}}{r_{\rm AB}^{l_{\rm A}+l_{\rm B}+1}}\right)=\frac{1}{r_{\rm AB}^{l_{\rm A}+l_{\rm B}+1}}\partial_{i}\left(\hat{n}_{L_{\rm A}L_{\rm B}}\right)+\hat{n}_{L_{\rm A}L_{\rm B}}\partial_i\left(\frac{1}{r_{\rm AB}^{l_{\rm A}+l_{\rm B}+1}}\right)\, ,
\end{equation}
and using the relation given by Hartmann, Soffel \& Kioustelidis (1994)
\begin{equation}
\label{truc1}
\partial_{i}\left(\hat{n}_{L_{\rm A}L_{\rm B}}\right)=\frac{l_{\rm A}+l_{\rm B}+1}{r_{\rm AB}}n_i\hat{n}_{L_{\rm A}L_{\rm B}}-\frac{2l_{\rm A}+2l_{\rm B}+1}{r_{\rm AB}}\hat{n}_{iL_{\rm A}L_{\rm B}}\, ,
\end{equation}
we get
\begin{equation}
\partial_i\left(\frac{\hat{n}_{L_{\rm A}L_{\rm B}}}{r_{\rm AB}^{l_{\rm A}+l_{\rm B}+1}}\right)=-\frac{2l_{\rm A}+2l_{\rm B}+1}{r_{\rm AB}^{l_{\rm A}+l_{\rm B}+2}}\hat{n}_{iL_{\rm A}L_{\rm B}}\, .\label{tayo2}
\end{equation}
Inserting Eqs. (\ref{derivdefou}) and (\ref{tayo2}) into Eq. (\ref{general2}), using Eq. (\ref{Ylm2STF}) with $\vec{z}_A-\vec{z}_B\equiv{\vec r}_{\rm AB}\equiv(r_{\rm AB},\theta_{\rm AB},\varphi_{\rm AB})$, we finally get the expression of the tidal potential for $\vert r\vert \le R_{\rm A}$
\begin{eqnarray}
\lefteqn{V^A_{\rm{T}}\left(t,\vec r,\vec r_{\rm AB}\right)=G\sum_{B\ne A}\left\lbrace \underbrace{\sum_{l_{\rm A}=2}^{\infty}\sum_{m_{\rm A}=-l_{\rm A}}^{l_{\rm A}}\sum_{l_{\rm B}=0}^{\infty}\sum_{l_{\rm B}=-m_{\rm B}}^{m_{\rm B}}(-1)^{l_{\rm A}}\frac{4\pi r^{l_{\rm A}}}{2l_{\rm A}+1}M^B_{l_{\rm B},m_{\rm B}}\gamma_{l_{\rm B},m_{\rm B}}^{l_{\rm A},m_{\rm A}}Y_{l_{\rm A},m_{\rm A}}^{*}(\theta,\varphi)\frac{Y_{l_{\rm A}+l_{\rm B},m_{\rm A}+m_{\rm B}}(\theta_{\rm AB},\varphi_{\rm AB})}{r_{\rm AB}^{l_{\rm A}+l_{\rm B}+1}}}_{\rm I}\right.}\nonumber\\
& &{\left.-\underbrace{\frac{1}{M_{\rm A}}\!\sum_{m_{\rm A}=-1}^{1}\sum_{l_{\rm A}^{'}=1}^\infty\sum_{m_{\rm A}^{'}=-l_{\rm A}^{'}}^{l_{\rm A}^{'}}\sum_{l_{\rm B}=0}^{\infty}\sum_{m_{\rm B}=-l_{\rm B}}^{l_{\rm B}}\!(-1)^{l_{\rm A}^{'}+1}\!\frac{4\pi}{3}r\!\left(2l_{\rm A}^{'}+2l_{\rm B}+1\right)\!M^{\rm A}_{l_{\rm A}^{'},m_{\rm A}^{'}}\!M^B_{l_{\rm B},m_{\rm B}}\!\gamma_{l_{\rm B},m_{\rm B}}^{l^{'}_{\rm A},m^{'}_{\rm A}}\!Y_{1,m_{\rm A}}^{*}(\theta,\varphi)\gamma_{l^{'}_{\rm A}+l_{\rm B},m^{'}_{\rm A}+m_{\rm B}}^{1,m_{\rm A}}\!\!\frac{Y_{l^{'}_{\rm A}+l_{\rm B}+1,m^{'}_{\rm A}+m_{\rm B}+m_{\rm A}}(\theta_{\rm AB},\varphi_{\rm AB})}{r_{\rm AB}^{l^{'}_{\rm A}+l_{\rm B}+2}}}_{\rm I\!I}\right\}}\, .\nonumber\\
\label{UTfinal}
\end{eqnarray}

The respective physical meanings of terms I and ${\rm I\!I}$ are clearly identified. Term I corresponds to the gravitational interaction of B with A, while term ${\rm I\!I}$ is the acceleration responsible for the movement of the center of mass of A. In the case of a ponctual mass perturber B, we recall that we get (see for example Melchior 1971):  
\begin{equation}
V_{\rm T}^{\rm A}(t,{\vec r},{\vec r}_{\rm AB})=V^{\rm B}\left(t,{\vec r},{\vec r}_{\rm AB}\right)-V_{\rm orb}^{\rm A}\left(t,{\vec r},{\vec r}_{\rm AB}\right)
\end{equation}
where
\begin{equation}
V^{\rm B}\left(t,{\vec r},{\vec r}_{\rm AB}\right)=G\frac{M_{\rm B}}{\vert{\vec r}-{\vec r}_{\rm AB}\vert}
\quad\hbox{and}\quad
V_{\rm orb}^{\rm A}\left(t,{\vec r},{\vec r}_{\rm AB}\right)=G\frac{M_{\rm B}}{r_{\rm AB}}\left(1+\frac{{\vec r}_{\rm AB}\cdot{\vec r}}{r_{\rm AB}^{2}}\right),
\end{equation}
$M_{\rm B}$ being its mass.\\
Eq. (\ref{UTfinal}) exactly corresponds to the Eq. (3.25) given in Hartmann, Soffel \& Kioustelidis (1994) that we recall here:
\begin{eqnarray}
V_{\rm T}^{\rm A}\left(t,\vec r,\vec r_{\rm AB}\right)&=&G\sum_{l_{\rm A}=2}^{\infty}\sum_{m_{\rm A}=-l_{\rm A}}^{l_{\rm A}}\sum_{l_{\rm B}=0}^{\infty}\sum_{m_{\rm B}=-l_{\rm B}}^{l_{\rm B}}\left(-1\right)^{l_{\rm A}}\frac{4\pi}{2l_{\rm A}+1}\gamma_{l_{\rm B},m_{\rm B}}^{l_{\rm A},m_{\rm A}}M_{l_{\rm B},m_{\rm B}}^{B}
r^{l_{\rm A}}Y_{l_{\rm A},m_{\rm A}}^{*}\left(\theta,\varphi\right)\frac{Y_{l_{\rm A}+l_{\rm B},m_{\rm A}+m_{\rm B}}\left(\theta_{\rm AB},\varphi_{\rm AB}\right)}{r_{\rm AB}^{l_{\rm A}+l_{\rm B}+1}}\nonumber\\
& &-\frac{G}{M_{\rm A}}\sum_{l_{\rm A}=1}^{\infty}\sum_{m_{\rm A}=-l_{\rm A}}^{l_{\rm A}}\sum_{l_{\rm B}=0}^{\infty}\sum_{m_{\rm B}=-l_{\rm B}}^{l_{\rm B}}\left(-1\right)^{l_{\rm A}}\gamma_{l_{\rm B},m_{\rm B}}^{l_{\rm A},m_{\rm A}}M_{l_{\rm A},m_{\rm A}}^{\rm A}M_{l_{\rm B},m_{\rm B}}^{\rm B}
X^{i}\partial_{i}\left[\frac{Y_{l_{\rm A}+l_{\rm B},m_{\rm A}+m_{\rm B}}\left(\theta_{\rm AB},\varphi_{\rm AB}\right)}{r_{\rm AB}^{l_{\rm A}+l_{\rm B}+1}}\right].
\end{eqnarray}
It is then recast in its general spectral form, using that $V_{\rm T}^{\rm A}$ is real and expanding it in the spherical harmonics for $\vert r \vert\le R_{\rm A}$:
\begin{equation}
V_{\rm T}^{\rm A}\left(t,\vec r,\vec r_{\rm AB}\right)=\sum_{{\rm B}\ne{\rm A}}G\sum_{l_{\rm A}=0}^{\infty}\sum_{m_{\rm A}=-l_{\rm A}}^{l_{\rm A}}\left[A_{{\rm I};l_{\rm A},m_{\rm A}}\left(t,{\vec r}_{\rm AB}\right)+A_{{\rm I\!I};l_{\rm A},m_{\rm A}}\left(t,{\vec r}_{\rm AB}\right)\right]r^{l_{\rm A}}Y_{l_{\rm A},m_{\rm A}}\left(\theta,\varphi\right),
\label{UTS}
\end{equation}
where the coefficients $A_{{\rm I};l_{\rm A},m_{\rm A}}$ and $A_{{\rm I\!I};l_{\rm A},m_{\rm A}}$ are respectively given by:
\begin{equation}
A_{{\rm I};l_{\rm A},m_{\rm A}}=\left(-1\right)^{l_{\rm A}}\frac{4\pi}{2l_{\rm A}+1}\left(1-\delta_{l_{\rm A},0}\right)\left(1-\delta_{l_{\rm A},1}\right)
\sum_{l_{\rm B}=0}^{\infty}\sum_{m_{\rm B}=-l_{\rm B}}^{l_{\rm B}}\gamma_{l_{\rm B},m_{\rm B}}^{l_{\rm A},m_{\rm A}}\left(M^{\rm B}_{l_{\rm B},m_{\rm B}}\right)^{*}\frac{Y_{l_{\rm A}+l_{\rm B},m_{\rm A}+m_{\rm B}}^{*}\left(\theta_{\rm AB},\varphi_{\rm AB}\right)}{r_{\rm AB}^{l_{\rm A}+l_{\rm B}+1}}
\label{A1}
\end{equation}
and
\begin{eqnarray}
A_{{\rm I\!I}; l_{\rm A},m_{\rm A}}&&=-\frac{1}{M_{\rm A}}\frac{4\pi}{3}\delta_{l_{\rm A},1}\sum_{l^{'}_{\rm A}=1}^{\infty}\sum_{m^{'}_{\rm A}=-l^{'}_{\rm A}}^{l^{'}_{\rm A}}\sum_{l_{\rm B}=0}^{\infty}\sum_{m_{\rm B}=-l_{\rm B}}^{l_{\rm B}}\left(-1\right)^{l^{'}_{\rm A}+1}\left(2l^{'}_{\rm A}+2l_{\rm B}+1\right)\gamma_{l_{\rm B},m_{\rm B}}^{l^{'}_{\rm A},m^{'}_{\rm A}}\left(M^{\rm A}_{l^{'}_{\rm A},m^{'}_{\rm A}}\right)^{*}\left(M^{\rm B}_{l_{\rm B},m_{\rm B}}\right)^{*}\gamma^{1,m_{\rm A}}_{l^{'}_{\rm A}+l_{\rm B},m^{'}_{\rm A}+m_{\rm B}}\nonumber\\
& &\times\frac{Y^{*}_{l^{'}_{\rm A}+l_{\rm B}+1, m^{'}_{\rm A}+m_{\rm B}+m_{\rm A}}\left(\theta_{\rm AB},\varphi_{\rm AB}\right)}{r_{\rm AB}^{l^{'}_{\rm A}+l_{\rm B}+2}}.
\label{A2}
\end{eqnarray}

\begin{figure}[h!]
\centering
\resizebox{12cm}{!}{\includegraphics{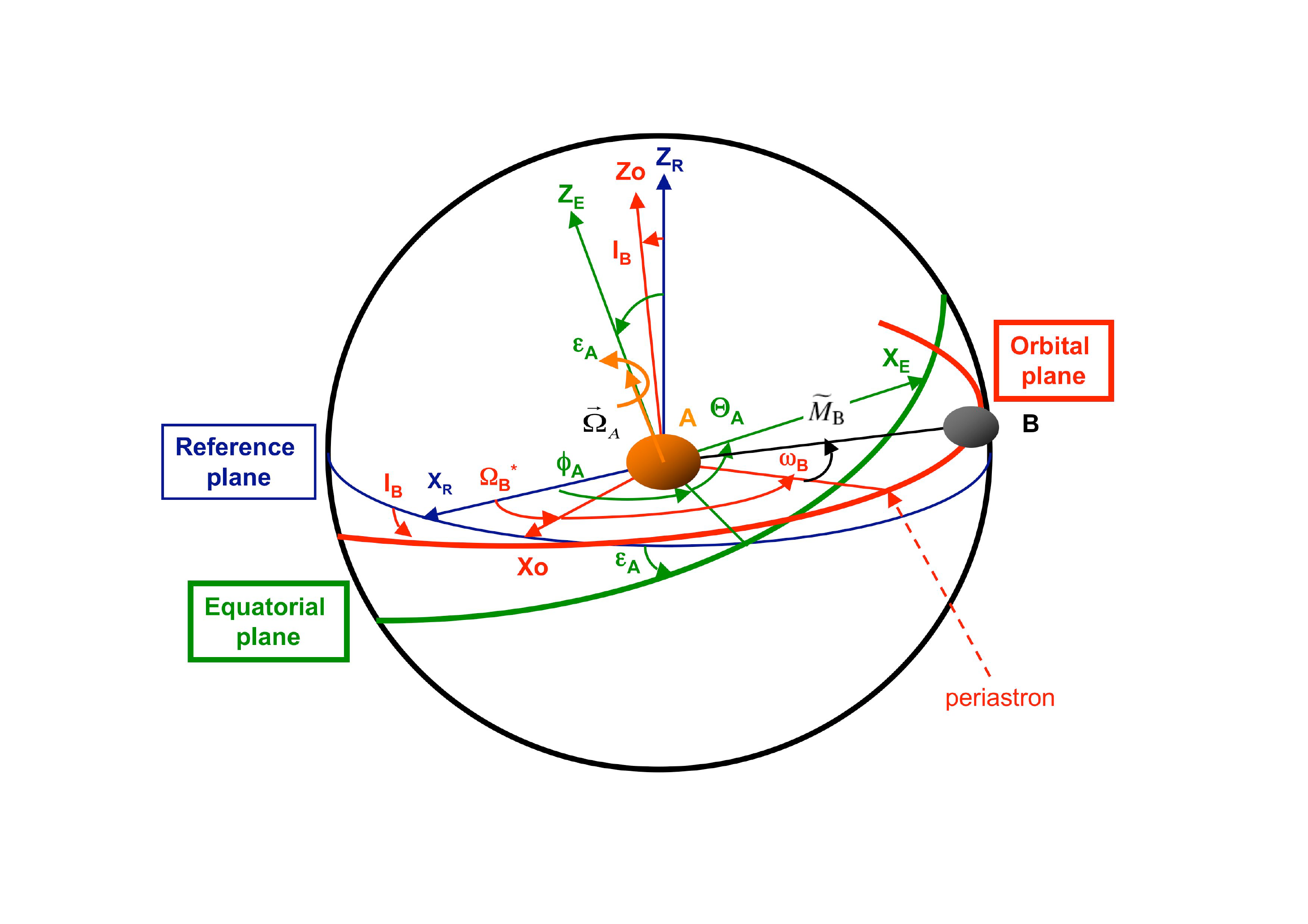}}
\caption{Inertial Reference, Orbital and Equatorial rotating frames (${\mathcal R}_{\rm R}$, ${\mathcal R}_{\rm O}$ and ${\mathcal R}_{E;{\rm T}}$) and associated Euler's angles of orientation.}
\label{fReferentiel}
\end{figure}

%In the same way as Zahn (1966), we define the tides as follow: it is related to all phenomena due to the time dependent part of the perturbation exerted by the companion, the static part being filtered (in other words, the tidal frequency $\sigma$ given in Eq. (\ref{Tfrequency}) is non zero).\\ 
The more general form of the tidal potential being derived, we now express $A_{{\rm I};l_{\rm A},m_{\rm A}}$ and $A_{{\rm I\!I}; l_{\rm A},m_{\rm A}}$ as a function of the Keplerian orbit elements of the perturber B.\\
Here, we take into account the relative inclinations of the spin of each body with respect to the orbital plane. It is then necessary to define three reference frames, represented on Fig. (\ref{fReferentiel}), all centered on the center of mass of the considered body A, $O_{\rm A}$:
\begin{itemize}
\item an inertial frame $\mathcal{R}_{\rm R}:\left\{O_{\rm A},{\bf X}_{\rm R},{\bf Y}_{\rm R},{\bf Z}_{\rm R}\right\}$, time independent, with ${\bf Z}_{\rm R}$ in the direction of the total angular momentum of the whole system ${\vec L}_{\rm Total}={\vec L}_{\rm Orbital}+{\vec L}_{\rm Body A}+\sum_{k}{\vec L}_{{\rm Body B}_{k}}$ which is a first integral (we are studying here the two bodies interaction between A and each potential perturber ${\rm B}_{k}$ with $k\in\left[\!\left[1,N\right]\!\right]$).\\
\item an orbital frame $\mathcal{R}_{\rm O}:\left\{O_{\rm A},{\bf X}_{\rm O},{\bf Y}_{\rm O},{\bf Z}_{\rm O}\right\}$. We define here three Euler angles to link this frame  to $\mathcal{R}_{\rm R}:\left\{O_{\rm A},{\bf X}_{\rm R},{\bf Y}_{\rm R},{\bf Z}_{\rm R}\right\}$:
\begin{itemize}
\item $I_{\rm B}$, the inclination of the orbital frame with respect to $\left(O_{\rm A},{\bf X}_{\rm R},{\bf Y}_{\rm R}\right)$;
\item $\omega_{\rm B}$, the argument of the pericenter;
\item $\Omega^{*}_{\rm B}$, the longitude of the ascending node.
\end{itemize}
Let us finally define the last three quantities associated to the elliptic elements of body B's center of mass: $a_{\rm B}$, the semi major axis, $e_{\rm B}$, the eccentricity and ${\widetilde M}_{\rm B}$, the mean anomaly with ${\widetilde M}_{\rm B}\approx n_{\rm B}t$, $n_{\rm B}$ being the mean motion.\\
\item a spin equatorial frame $\mathcal{R}_{{\rm E;T}}:\left\{O_{\rm A},{\bf X}_{\rm E},{\bf Y}_{\rm E},{\bf Z}_{\rm E}\right\}$. This frame is rotating with the angular velocity, $\Omega_{\rm A}$. This frame is linked to $\mathcal{R}_{\rm R}:\left\{O_{\rm A},{\bf X}_{\rm R},{\bf Y}_{\rm R},{\bf Z}_{\rm R}\right\}$ by three Euler angles:\\
\begin{itemize}
\item $\varepsilon_{\rm A}$, the obliquity, {\it i.e.} the inclination of the equatorial plane with respect to the reference plane $\left(O_{\rm A},{\bf X}_{\rm R},{\bf Y}_{\rm R}\right)$;
\item $\Theta_{\rm A}$, the mean sideral angle where $\Theta_{\rm A}={\rm d}\Omega_{\rm A}/{\rm d}t$. This is the angle between the minimal axis of inertia and the straight line due to the intersection of the planes $\left(O_{\rm A},{\bf X}_{\rm E},{\bf Y}_{\rm E}\right)$ and $\left(O_{\rm A},{\bf X}_{\rm R},{\bf Y}_{\rm R}\right)$.
\item $\phi_{\rm A}$, the general precession angle.\\
\end{itemize}
\end{itemize}

The Kaula's transform is then used to explicitly express all the generic multipole expansion in spherical harmonics in term of keplerian elements. Using the results derived by Kaula (1962), the following identity is obtained: 
\begin{equation}
\frac{Y_{l,m}\left(\theta_{\rm AB},\varphi_{\rm AB}\right)}{r_{\rm AB}^{l+1}}=\frac{1}{a_{\rm B}^{l+1}}\sum_{j=-l}^{l}\sum_{p=0}^{l}\sum_{q}\kappa_{l,j}d_{j,m}^{l}\left(\varepsilon_{\rm A}\right)F_{l,j,p}\left(I_{\rm B}\right)G_{l,p,q}\left(e_{\rm B}\right)\exp\left[i\Psi_{l,m,j,p,q}\right],
\label{Kaula1}
\end{equation}
where the $\kappa_{l,j}$ coefficients are given by:
\begin{equation}
\kappa_{l,j}=\sqrt{\frac{2l+1}{4\pi}\frac{\left(l-|j|\right)!}{\left(l+|j|\right)!}}.
\label{Kaula2}
\end{equation}
$d_{j,m}^{l}\left(\varepsilon_{\rm A}\right)$ is the obliquity function which is defined as follow for $j\ge m$:
\begin{equation}
d_{j,m}^{l}\left(\varepsilon_{\rm A}\right)=(-1)^{j-m}\left[\frac{\left(l+j\right)!\left(l-j\right)!}{\left(l+m\right)!\left(l-m\right)!}\right]^{1\over2}\left[\cos\left(\frac{\varepsilon_{\rm A}}{2}\right)\right]^{j+m}\left[\sin\left(\frac{\varepsilon_{\rm A}}{2}\right)\right]^{j-m}P_{l-j}^{\left(j-m,j+m\right)}\left(\cos\varepsilon_{\rm A}\right),
%\sum_{n=\max\left(0,l-j\right)}^{\min\left(l+m,l-m\right)}\Bigg\lbrace\left(-1\right)^{n}\begin{pmatrix}
%l+m\\
%n\\
%\end{pmatrix}
%\begin{pmatrix}
%l-m\\
%l-j-n\\
%\end{pmatrix}\left(\cos\frac{\varepsilon_{\rm A}}{2}\right)^{2l-j+m-2n}\left(\sin\frac{\varepsilon_{\rm A}}{2}\right)^{j-m+2n}\Bigg\rbrace,
\label{obfunc}
\end{equation}
the $P_{l}^{\left(\alpha,\,\beta\right)}\left(x\right)$ being the Jacobi polynomials (cf. Abramowitz \& Stegun, 1972). The value of the function for indices $j$ which do not verify $j\ge m$ are deduced from:
\begin{equation}
d_{j,m}^{l}\left(\pi+\varepsilon_{\rm A}\right)=\left(-1\right)^{l-j}d_{-j,m}^{l}\left(\varepsilon_{\rm A}\right)
\end{equation}
or from their symmetry properties:
\begin{equation}
d_{j,m}^{l}\left(\varepsilon_{\rm A}\right)=(-1)^{j-m}d_{-j,-m}^{l}\left(\varepsilon_{\rm A}\right)=d_{m,j}^{l}\left(-\varepsilon_{\rm A}\right).
\end{equation}
On the other hand, one should note that: $d_{j,m}^{l}\left(0\right)=\delta_{jm}$.\\
\begin{table}[h!]              
\centering          
\fbox{\begin{tabular}{c c c l l}  
\hline      
\multicolumn{2}{c}{} & \vline &\multicolumn{1}{c}{}\\
 $j$ & $m$ & \vline & $d^{2}_{j,m}\left(\varepsilon\right)$ \\              
\multicolumn{2}{c}{} & \vline &\multicolumn{1}{c}{}\\  
\hline    
 & & \vline & \\   
 2 & 2 & \vline & $\left(\cos\frac{\varepsilon}{2}\right)^{4}$\\
 & & \vline &\\             
 2 & 1 & \vline & $-2\left(\cos\frac{\varepsilon}{2}\right)^{3}\left(\sin\frac{\varepsilon}{2}\right)$\\
 & & \vline & \\   
 2 & 0 & \vline & $\sqrt{6}\left(\cos\frac{\varepsilon}{2}\right)^{2}\left(\sin\frac{\varepsilon}{2}\right)^{2}$\\
 & & \vline &\\             
 1 & 1 & \vline & $\left(\cos\frac{\varepsilon}{2}\right)^{4}-3\left(\cos\frac{\varepsilon}{2}\right)^{2}\left(\sin\frac{\varepsilon}{2}\right)^{2}$\\
 & & \vline &\\             
 1 & 0 & \vline & $-\sqrt{6}\cos\varepsilon\left(\cos\frac{\varepsilon}{2}\right)\left(\sin\frac{\varepsilon}{2}\right)$\\
 & & \vline &\\     
 0 & 0 & \vline & $1-6\left(\cos\frac{\varepsilon}{2}\right)^{2}\left(\sin\frac{\varepsilon}{2}\right)^{2}$\\
 & & \vline &\\    
\hline                  
\end{tabular}}
\caption{Values of the obliquity function $d_{j,m}^{l}\left(\varepsilon\right)$ in the case where $l=2$ and $j\ge m$ obtained from Eq. (\ref{obfunc}) (adapted from Yoder, 1995).}  
\label{dsjm}
\end{table}
The inclination function, $F_{l,j,p}\left(I_{\rm B}\right)$, is defined in a similar way:
\begin{equation}
F_{l,j,p}\left(I_{\rm B}\right)=\left(-1\right)^{p}\left[\frac{4\pi}{2l+1}\frac{\left(l+j\right)!}{\left(l-j\right)!}\right]^{1\over2}Y_{l,l-2p}\left(\frac{\pi}{2},0\right)d_{l-2p,j}^{l}\left(-I_{\rm B}\right),
\end{equation}
where
\begin{equation}
Y_{l,m}\left(\frac{\pi}{2},0\right)=\left[\frac{2l+1}{4\pi}\right]^{1\over2}\frac{\left[\left(l-m\right)!\left(l+m\right)!\right]^{1\over2}}{2^{l}\left[\left(l-m\right)/2\right]!\left[\left(l+m\right)/2\right]!}\cos\left[\left(l-m\right)\frac{\pi}{2}\right];
\end{equation}
moreover, the following symmetry property is verified:
\begin{equation}
F_{l,-j,p}\left(I_{\rm B}\right)=\left[\left(-1\right)^{l-j}\frac{\left(l-j\right)!}{\left(l+j\right)!}\right]F_{l,j,p}\left(I_{\rm B}\right).
\label{Fsym}
\end{equation}
The usual value of these functions are given in Tab. (\ref{Fljp}).\\
\begin{table}[h!]            
\centering          
\fbox{\begin{tabular}{c c c c l l}  
\hline      
\multicolumn{3}{c}{} & \vline &\multicolumn{1}{c}{}\\
 $l$ & $j$ & $p$ & \vline & $F_{l,j,p}\left(I\right)$ \\              
\multicolumn{3}{c}{} & \vline &\multicolumn{1}{c}{}\\  
\hline    
 & & & \vline & \\   
 2 & 0 & 0 & \vline & $\frac{3}{8}\sin^{2}I$\\
 & & & \vline &\\             
 2 & 0 & 1 & \vline & $-\frac{3}{4}\sin^{2}I+\frac{1}{2}$\\
 & & & \vline & \\   
 2 & 0 & 2 & \vline & $\frac{3}{8}\sin^{2}I$\\
 & & & \vline &\\             
 2 & 1 & 0 & \vline & $\frac{3}{4}\sin I\left(1+\cos I\right)$\\
 & & & \vline &\\             
 2 & 1 & 1 & \vline & $-\frac{3}{2}\sin I\cos I$\\
 & & & \vline &\\     
 2 & 1 & 2 & \vline & $-\frac{3}{4}\sin I\left(1-\cos I\right)$\\
 & & & \vline &\\    
 2 & 2 & 0 & \vline & $\frac{3}{4}\left(1+\cos I\right)^{2}$\\
 & & & \vline &\\     
 2 & 2 & 1 & \vline & $\frac{3}{2}\sin^{2}I$\\
 & & & \vline &\\     
 2 & 2 & 2 & \vline & $\frac{3}{4}\left(1-\cos I\right)^{2}$\\
 & & & \vline &\\     
\hline                  
\end{tabular}}
\caption{Values of the inclination function $F_{l,j,p}\left(I\right)$ in the case where $l=2$. Values for $j<0$ can be deduced from Eq. (\ref{Fsym}) (adapted from Lambeck, 1980). }  
\label{Fljp}
\end{table}
The eccentricity functions $G_{l,p,q}\left(e_{\rm B}\right)$, are polynomial functions having $e_{\rm B}^{q}$ for argument (see Kaula, 1962 and Laskar, 2005 for their detailed properties). Their values for usual sets $\left\{l,p,q\right\}$ are given in Tab. (\ref{Gspqe}). In the case of weakly eccentric orbits, the summation over a small number of values for $q$ is sufficient ($q\in\left[\!\left[-2,2\right]\!\right]$). More details can be found in the appendix of Yoder (1995).\\
\begin{table}[h!]          
\centering          
\fbox{\begin{tabular}{c c c c c c c l l}  
\hline      
\multicolumn{3}{c}{} & \vline &\multicolumn{3}{c}{} & \vline &\multicolumn{1}{c}{}\\
 $l$ & $p$ & $q$ & \vline & $l$ & $p$ & $q$ & \vline & $G_{l,p,q}\left(e\right)$ \\              
\multicolumn{3}{c}{} & \vline & \multicolumn{3}{c}{} & \vline &\multicolumn{1}{c}{}\\  
\hline    
 & & & \vline & & & & \vline & \\   
 2 & 0 & -2& \vline & 2 & 2 &  2 & \vline & 0\\
 & & & \vline & & & & \vline &\\             
 2 & 0 & -1 & \vline & 2 & 2 & 1 & \vline & $-\frac{1}{2}e+\cdot\cdot\cdot$\\
 & & & \vline & & & & \vline & \\   
 2 & 0 & 0 & \vline & 2 & 2 &  0 & \vline & $1-\frac{5}{2}e^{2}+\cdot\cdot\cdot$\\
 & & & \vline & & & & \vline &\\             
 2 & 0 & 1 & \vline & 2 & 2 & -1 & \vline & $\frac{7}{2}e+\cdot\cdot\cdot$\\
 & & & \vline & & & & \vline &\\             
 2 & 0 & 2 & \vline & 2 & 2 & -2 & \vline & $\frac{17}{2}e^{2}+\cdot\cdot\cdot$\\
 & & & \vline & & & & \vline &\\     
 2 & 1 & -2 & \vline & 2 & 1 & 2 & \vline & $\frac{9}{4}e^{2}+\cdot\cdot\cdot$\\
 & & & \vline & & & & \vline &\\    
 2 & 1 & -1 & \vline & 2 & 1 & 1 & \vline & $\frac{3}{2}e+\cdot\cdot\cdot$\\
 & & & \vline & & & & \vline &\\     
 &  &  & \vline & 2 & 1 & 0 & \vline & $\left(1-e^{2}\right)^{-3/2}$\\
 & & & \vline & & & & \vline &\\     
\hline                  
\end{tabular}}
\caption{Values of the eccentricity function $G_{l,p,q}\left(e\right)$ in the case where $l=2$ (adapted from Lambeck, 1980). {Note that several combinations $\left\{l,p,q\right\}$ have the same $G_{l,p,q}$ value.}}
\label{Gspqe}
\end{table}
Finally, the phase argument is given by:
\begin{equation}
\Psi_{l,m,j,p,q}=\left(l-2p+q\right){\widetilde M}_{\rm B}+\Phi_{l,m,j,p,q}\left(\omega_{\rm B},\Omega^{*}_{\rm B},\Theta_{\rm A},\phi_{\rm A}\right)
\label{Kaula3}
\end{equation}
where
\begin{equation}
\Phi_{l,m,j,p,q}=\left(l-2p\right)\omega_{\rm B}+j\left(\Omega^{*}_{\rm B}-\phi_{\rm A}\right)-m\Theta_{\rm A}+\left(l-m\right)\frac{\pi}{2}.
\end{equation}
This can be also written as:
\begin{equation}
\Psi_{l,m,j,p,q}=\sigma_{l,m,p,q}\left(n_{\rm B},\Omega_{\rm A}\right)t+\psi_{l,m,j,p,q}\left(\omega_{\rm B},\Omega^{*}_{\rm B},\phi_{\rm A}\right),
\end{equation}
where we have defined the tidal frequency:
\begin{equation}
\sigma_{l,m,p,q}=(l-2p+q)n_{\rm B}-m\Omega_{\rm A}
\label{Tfrequency}
\end{equation}
and
\begin{equation}
\psi_{l,m,j,p,q}=\left(l-2p\right)\omega_{\rm B}+j\left(\Omega^{*}_{\rm B}-\phi_{\rm A}\right)+\left(l-m\right)\frac{\pi}{2}.
\end{equation}
The Kaula's transform allows us to express each function of $\vec r_{\rm AB}$, {\it i. e.} of $\left(r_{\rm AB},\theta_{\rm AB},\varphi_{\rm AB}\right)$, as a function of the Keplerian relative orbital elements of B in the A-frame. Applying Eq. (\ref{Kaula1}) to $A_{{\rm I};l_{\rm A},m_{\rm A}}$ and $A_{{\rm I\!I}; l_{\rm A},m_{\rm A}}$ respectively given in Eq. (\ref{A1}) and in Eq. (\ref{A2}), we get:
\begin{eqnarray}
A_{{\rm I};l_{\rm A},m_{\rm A}}&=&\left(-1\right)^{l_{\rm A}}\frac{4\pi}{2l_{\rm A}+1}\left(1-\delta_{l_{\rm A},0}\right)\left(1-\delta_{l_{\rm A},1}\right)
\sum_{l_{\rm B}=0}^{\infty}\sum_{m_{\rm B}=-l_{\rm B}}^{l_{\rm B}}\gamma_{l_{\rm B},m_{\rm B}}^{l_{\rm A},m_{\rm A}}|M_{l_{\rm B},m_{\rm B}}^{\rm B}|\exp\left[-i\delta M_{l_{\rm B},m_{\rm B}}^{\rm B}\right]\nonumber\\
& &\times\frac{1}{a_{\rm B}^{l_{\rm A}+l_{\rm B}+1}}\sum_{j=-\left(l_{\rm A}+l_{\rm B}\right)}^{l_{\rm A}+l_{\rm B}}\sum_{p=0}^{l_{\rm A}+l_{\rm B}}\sum_{q}\kappa_{l_{\rm A}+l_{\rm B},j}d_{j,m_{\rm A}+m_{\rm B}}^{l_{\rm A}+l_{\rm B}}\left(\varepsilon_{\rm A}\right)
F_{l_{\rm A}+l_{\rm B},j,p}\left(I_{\rm B}\right) G_{l_{\rm A}+l_{\rm B},p,q}\left(e_{\rm B}\right) \exp\left[-i\Psi_{l_{\rm A}+l_{\rm B},m_{\rm A}+m_{\rm B},j,p,q}\right]\nonumber\\
\label{A3}
\end{eqnarray}
and
\begin{eqnarray}
A_{{\rm I\!I}; l_{\rm A},m_{\rm A}}&=&-\frac{1}{M_{\rm A}}\frac{4\pi}{3}\delta_{l_{\rm A},1}\sum_{l^{'}_{\rm A}=1}^{\infty}\sum_{m^{'}_{\rm A}=-l^{'}_{\rm A}}^{l^{'}_{\rm A}}\sum_{l_{\rm B}=0}^{\infty}\sum_{m_{\rm B}=-l_{\rm B}}^{l_{\rm B}}\left(-1\right)^{l^{'}_{\rm A}+1}
\left(2l^{'}_{\rm A}+2l_{\rm B}+1\right)\gamma_{l_{\rm B},m_{\rm B}}^{l^{'}_{\rm A},m^{'}_{\rm A}}|M_{l^{'}_{\rm A},m^{'}_{\rm A}}^{\rm A}|\exp\left[-i\delta M_{l^{'}_{\rm A},m^{'}_{\rm A}}^{\rm A}\right]
|M_{l_{\rm B},m_{\rm B}}^{\rm B}|\exp\left[-i\delta M_{l_{\rm B},m_{\rm B}}^{\rm B}\right]\nonumber\\
& &\times\gamma^{1,m_{\rm A}}_{l^{'}_{\rm A}+l_{\rm B},m^{'}_{\rm A}+m_{\rm B}}
\frac{1}{a_{\rm B}^{l^{'}_{\rm A}+l_{\rm B}+2}}\sum_{r=-\left(l^{'}_{\rm A}+l_{\rm B}+1\right)}^{l^{'}_{\rm A}+l_{\rm B}+1}\sum_{s=0}^{l^{'}_{\rm A}+l_{\rm B}+1}\sum_{u}\kappa_{l^{'}_{\rm A}+l_{\rm B}+1,r}
d_{r,m^{'}_{\rm A}+m_{\rm B}+m_{\rm A}}^{l^{'}_{\rm A}+l_{\rm B}+1}\left(\varepsilon_{\rm A}\right)F_{l^{'}_{\rm A}+l_{\rm B}+1,r,s}\left(I_{\rm B}\right)G_{l^{'}_{\rm A}+l_{\rm B}+1,s,u}\left(e_{\rm B}\right)\nonumber\\
& &\times\exp\left[-i\Psi_{l^{'}_{\rm A}+l_{\rm B}+1,m^{'}_{\rm A}+m_{\rm B}+m_{\rm A},r,s,u}\right],
\label{A4}
\end{eqnarray}
{where as in Eq. (\ref{GME}) $M_{l_{\rm B},m_{\rm B}}^{\rm B}=M_{l_{\rm B},m_{\rm B}}^{\rm S_{\rm B}}+M_{l_{\rm B},m_{\rm B}}^{\rm T_{\rm B}}$ and $M_{l_{\rm A},m_{\rm A}}^{\rm A}=M_{l_{\rm A},m_{\rm A}}^{\rm S_{\rm A}}+M_{l_{\rm A},m_{\rm A}}^{\rm T_{\rm A}}$.

Like in Zahn (1966-1977), the tidal potential can be splitted into two components. The first one,  $V_{{\rm T};1}^{\rm A}\left(\vec r,\vec r_{\rm AB}\right)$, is stationary ({\it i.e.} the tidal frequency vanishes: $\sigma=0$). It corresponds to the axisymmetric permanent deformation induced by B. In the case of a ponctual mass perturber and of a system where all the spins are aligned, Zahn (1966-1977) shown that $V_{{\rm T};1}^{\rm A}=-\frac{G M_{\rm B}}{a_{\rm B}^{3}}\frac{1}{2}\left(1-e_{\rm B}^2\right)^{-3/2}r^2P_{2}\left(\cos\theta\right)$. Then, the second component is the time dependent part of the perturbation, $V_{{\rm T};2}^{\rm A}\left(t,\vec r,\vec r_{\rm AB}\right)$, for which $\sigma\ne0$.}
\subsection{The two bodies interaction potential}
The mutual gravitational interaction potential\footnote{the denomination of $V_{\rm A-B}$ as a potential is not very pertinent since it has the dimension of the product of a mass by a potential. However, we keep it to stay coherent with Hartmann, Soffel, Kioustelidis (1994).} of two bodies A and B is defined as:
\begin{equation}
V_{{\rm A}-{\rm B}}\left(t,{\vec r}_{\rm AB}\right)=\int_{M_{\rm A}}V^{\rm B}\left(t,{\vec r},{\vec r}_{\rm AB}\right){\rm d}M_{\rm A}.
\end{equation}
Following Hartmann, Soffel, Kioustelidis (1994), its expansion on STF-tensors is given by
\begin{equation}
V_{\rm A-B}=G \sum_{l_{\rm A}=0}^{\infty}\sum_{l_{\rm B}=0}^{\infty}\frac{\left(-1\right)^{l_{\rm B}}}{l_{\rm A}!l_{\rm B}!}{\mathcal M}_{\rm A}^{L_{\rm A}}{\mathcal M}_{\rm B}^{L_{\rm B}}\partial^{\rm A}_{L_{\rm A}L_{\rm B}}\left(\frac{1}{r_{\rm AB}}\right).
\end{equation}
Using once again Eqs. (\ref{composition1}) and (\ref{MlmSTF}), we get:
\begin{equation}
V_{\rm A-B}=G \sum_{l_{\rm A}=0}^{\infty}\sum_{m_{\rm A}=-l_{\rm A}}^{l_{\rm A}}\sum_{l_{\rm B}=0}^{\infty}\sum_{m_{\rm B}=-l_{\rm B}}^{l_{\rm B}}\left\{M_{l_{\rm A},m_{\rm A}}^{\rm A}M_{l_{\rm B},m_{\rm B}}^{\rm B}\left(-1\right)^{l_{\rm A}}\gamma_{l_{\rm B},m_{\rm B}}^{l_{\rm A},m_{\rm A}}\frac{Y_{l_{\rm A}+l_{\rm B},m_{\rm A}+m_{\rm B}}\left(\theta_{\rm AB},\varphi_{\rm AB}\right)}{r_{\rm AB}^{l_{\rm A}+l_{\rm B}+1}}\right\}.
\label{Uint}
\end{equation}
Finally, using the Kaula transformation given in Eqs. (\ref{Kaula1}), (\ref{Kaula2}) and (\ref{Kaula3}) as previously done for $V_{\rm T}^{\rm A}$, $V_{\rm A-\rm B}$ is expressed as a function of the obliquity, $\varepsilon_{\rm A}$, and of the Keplerian orbital elements of B: $a_{\rm B}$, $e_{\rm B}$ and $I_{\rm B}$:
\begin{eqnarray}
V_{{\rm A}-{\rm B}}&=&G\sum_{l_{\rm A}=0}^{\infty}\sum_{m_{\rm A}=-l_{\rm A}}^{l_{\rm A}}\sum_{l_{\rm B}=0}^{\infty}\sum_{m_{\rm B}=-l_{\rm B}}^{l_{\rm B}}\left\{M_{l_{\rm A},m_{\rm A}}^{\rm A}M_{l_{\rm B},m_{\rm B}}^{\rm B}\left(-1\right)^{l_{\rm A}}\gamma_{l_{\rm B},m_{\rm B}}^{l_{\rm A},m_{\rm A}}\right.\nonumber\\
& &\times{\left.\frac{1}{a_{\rm B}^{l_{\rm A}+l_{\rm B}+1}}\sum_{v=-\left(l_{\rm A}+l_{\rm B}\right)}^{\left(l_{\rm A}+l_{\rm B}\right)}\sum_{w=0}^{l_{\rm A}+l_{\rm B}}\sum_{b}\kappa_{l_{\rm A}+l_{\rm B},v}
d_{v,m_{\rm A}+m_{\rm B}}^{l_{\rm A}+l_{\rm B}}\left(\varepsilon_{\rm A}\right)F_{l_{\rm A}+l_{\rm B},v,w}\left(I_{\rm B}\right)G_{l_{\rm A}+l_{\rm B},w,b}\left(e_{\rm B}\right)\exp\left[i\Psi_{l_{\rm A}+l_{\rm B},m_{\rm A}+m_{\rm B},v,w,b}\right]\right\}}.
\end{eqnarray}
This interaction potential contains all multipole-multipole couplings. It is used in section 3.2 to compute the disturbing function aimed to study the dynamics of an extended body in gravitational interaction with A.\\
\par Since all type of gravitational potentials have been examined, we now study the dynamics of a system of extended bodies. 

\section{Equations of motion}

\subsection{External gravitational potential of a tidally perturbed body}

The goal of this section is to derive the external gravitational potential of a tidally perturbed extended body A by an extended body B. This potential is the sum of the structural self-gravitational potential of A, $V_{\rm S}^{\rm A}\left(t,\vec r\right)$, and of ${\widetilde V}_{\rm T}^{\rm A}\left(t,\vec r,\vec r_{\rm AB}\right)$, the tidally induced gravitational potential corresponding to the response of A to the perturbing potential $V_{\rm T}^{\rm A}\left(t,\vec r,\vec r_{\rm AB}\right)$:
\begin{equation}
V_{\rm ext}^{\rm A}\left(t,\vec r,\vec r_{\rm AB}\right)=V_{\rm S}^{\rm A}\left(t,\vec r\right)+{\widetilde V}_{\rm T}^{\rm A}\left(t,\vec r,\vec r_{\rm AB}\right)
\end{equation}
with the following definition for $V_{\rm S}$:
\begin{equation}
V_{\rm S}^{\rm A}\left(t,\vec r\right)=G\sum_{l_{\rm A}=0}^{\infty}\sum_{l_{\rm A}=-m_{\rm A}}^{m_{\rm A}}M_{l_{\rm A},m_{\rm A}}^{{\rm S}_{\rm A}}\frac{Y_{l_{\rm A},m_{\rm A}}\left(\theta,\varphi\right)}{r^{l_{\rm A}+1}}
\end{equation}
where $M_{l_{\rm A},m_{\rm A}}^{{\rm S}_{\rm A}}$ are the multipole moments of A in the case where it is not tidally perturbed by any other body, in other words, in the case it is isolated. By definition the external gravitational potential is harmonic; therefore $V_{\rm ext}^{\rm A}\left(t,\vec r,\vec r_{\rm AB}\right)$ verifies the Laplace equation:
\begin{equation}
\nabla^{2}V^{\rm A}_{\rm ext}\left(t,\vec r,\vec r_{\rm AB}\right)=0\quad\hbox{if}\quad |\vec r|\ge R_{\rm A},
\end{equation}
that directly leads to the same equation for ${\widetilde V}_{\rm T}^{\rm A}\left(t,\vec r,\vec r_{\rm AB}\right)$:
\begin{equation}
\nabla^{2}{\widetilde V}_{\rm T}^{\rm A}\left(t,\vec r,\vec r_{\rm AB}\right)=0\quad\hbox{if}\quad |\vec r|\ge R_{\rm A}.
\end{equation}
Following Lambeck (1980), N\'eron  de Surgy (1996), N\'eron  de Surgy \& Laskar (1997) and Correia \& Laskar (2003a, 2003b), we use the classical Love numbers, $k^{\rm A}_{l_{\rm A}}$, which allow us to characterize the response of the body A to the tidal perturbation. The boundary conditions for ${\widetilde V}_{\rm T}^{\rm A}\left(t,\vec r,\vec r_{\rm AB}\right)$ are:
\begin{equation}
\left\{
\begin{array}{l@{\quad}l}
{\widetilde V}_{\rm T}^{\rm A}\left(t,\vert\vec r\vert\rightarrow 0,\vec r_{\rm AB}\right)=0\\
{\widetilde V}_{\rm T}^{\rm A}\left(t,\vert\vec r\vert=R_{\rm A},\vec r_{\rm AB}\right)=\sum_{l_{\rm A}}k^{\rm A}_{l_{\rm A}}V_{l_{\rm A}}\left(t,\vert\vec r\vert=R_{\rm A},\vec r_{\rm AB}\right)
\end{array}
\right. ,
\label{BCVT}
\end{equation}
where $V_{l_{\rm A}}$ is the $l_{\rm A}^{\mbox{th}}$ spherical harmonic of $V_{\rm T}^{\rm A}\left(t,\vec r,\vec r_{\rm AB}\right)$. We also recall that using Eq. (\ref{UTS}) $V_{\rm T}^{\rm A}\left(t,\vec r,\vec r_{\rm AB}\right)$ has been expanded as follow for $\vert r\vert\le R_{\rm A}$:
\begin{equation}
V_{\rm T}^{\rm A}\left(t,\vec r,\vec r_{\rm AB}\right)=G\sum_{l_{\rm A},m_{\rm A}}\left[A_{{\rm I};l_{\rm A},m_{\rm A}}\left(t,\vec r_{\rm AB}\right)+A_{{\rm I\!I};l_{\rm A},m_{\rm A}}\left(t,\vec r_{\rm AB}\right)\right]r^{l_{\rm A}}Y_{l_{\rm A},m_{\rm A}}\left(\theta,\varphi\right).
\label{CD1}
\end{equation}
Using the well-known properties of the Laplace's equation, we search the solution for ${\widetilde V}_{\rm T}^{\rm A}$ when $\vert r\vert \ge R_{\rm A}$ of the form:
\begin{equation}
{\widetilde V}_{\rm T}^{\rm A}\left(t,\vec r,\vec r_{\rm AB}\right)=G\sum_{l_{\rm A}=0}^{\infty}\sum_{m_{\rm A}=-l_{\rm A}}^{l_{\rm A}}M^{{\rm T}_{\rm A}}_{l_{\rm A},m_{\rm A}}\left(t,\vec r_{\rm AB}\right)\frac{Y_{l_{\rm A},m_{\rm A}}\left(\theta,\varphi\right)}{{r}^{l_{\rm A}+1}}.
\label{CD2}
\end{equation}
Inserting Eqs. (\ref{CD1}) and (\ref{CD2}) into (\ref{BCVT}), the final solution of ${\widetilde V}_{\rm T}^{\rm A}$ is then derived with: 
\begin{equation}
M^{{\rm T}_{\rm A}}_{l_{\rm A},m_{\rm A}}=M^{{\rm T}_{\rm A};\rm I}_{l_{\rm A},m_{\rm A}}\left(t,\vec r_{\rm AB}\right)+M^{{\rm T}_{\rm A};\rm I\!I}_{l_{\rm A},m_{\rm A}}\left(t,\vec r_{\rm AB}\right)
\label{MTA}
\end{equation}
where $M^{{\rm T}_{\rm A};\rm I}_{l_{\rm A},m_{\rm A}}$ and $M^{{\rm T}_{\rm A};\rm I\!I}_{l_{\rm A},m_{\rm A}}$ are given by:
\begin{equation}
\left\{
\begin{array}{l@{\quad}l}
M^{{\rm T}_{\rm A};\rm I}_{l_{\rm A},m_{\rm A}}=k^{\rm A}_{l_{\rm A}}A_{{\rm I};l_{\rm A},m_{\rm A}}R_{\rm E}^{2l_{\rm A}+1}\\
M^{{\rm T}_{\rm A};\rm I\!I}_{l_{\rm A},m_{\rm A}}=k^{\rm A}_{l_{\rm A}}A_{{\rm I\!I};l_{\rm A},m_{\rm A}}R_{\rm E}^{2l_{\rm A}+1}
\end{array}
\right. .
\label{MTf}
\end{equation}

The response of the body A, which is described by the Love numbers, is the adiabatic one. However, it is well known that an elastic as well as a fluid body reacts to the tidal perturbation with a damping and a time delay which are due to the internal friction and diffusivities (in other words to the viscosity, $\nu$, and the thermal diffusivity, $K$, in a non-magnetic body). That allows us to transform the mechanical energy into thermal one which leads us to the dynamical evolution of the studied system (cf. Fig.(\ref{fMaree})). Therefore, we introduce a complex impedance, $Z_{{\rm T}_{\rm A}}\left(\nu,K;\Psi_{L}\right)$, with its associated argument, $\delta_{{\rm T}_{\rm A}}\left(\nu,K;\Psi_{L}\right)$:
\begin{equation}
Z_{{\rm T}_{\rm A}}\left(\nu,K;\Psi_{L}\right)=\vert Z_{{\rm T}_{\rm A}}\left(\nu,K;\Psi_{L}\right)\vert\exp\left[i \delta_{{\rm T}_{\rm A}}\left(\nu,K;\Psi_{L}\right)\right]
\end{equation}
which describes this damping. We thus substitute $ k^{\rm A}_{l_{\rm A}}\vert Z_{\rm T_{\rm A}}\vert \exp\left[i\delta_{{\rm T}_{\rm A}}\right]$ to $k^{\rm A}_{l_{\rm A}}$ in the Eq. (\ref{MTf}). $L$ corresponds to the indices of the considered tidal Fourier's mode\footnote{Note that each tidal Fourier's mode have its own dissipation rate as it as been shown by Zahn (1966-1977).}. The different modellings that can be adopted for $Z_{{\rm T}_{\rm A}}$ and $\delta_{{\rm T}_{\rm A}}$ are described in Alexander (1973), Zahn (1977) and Correia \& Laskar (2003a).\\

Using Eqs. (\ref{A3}) and (\ref{A4}), the expression of $M^{{\rm T}_{\rm A};\rm I}_{l_{\rm A},m_{\rm A}}$ and $M^{{\rm T}_{\rm A};\rm I\!I}_{l_{\rm A},m_{\rm A}}$ are obtained:
\begin{eqnarray}
\lefteqn{M^{{\rm T}_{\rm A};\rm I}_{l_{\rm A},m_{\rm A}}=\left(-1\right)^{l_{\rm A}}\frac{4\pi}{2l_{\rm A}+1}k_{l_{\rm A}}^{\rm A}R_{\rm A}^{2l_{\rm A}+1}\left(1-\delta_{l_{\rm A},0}\right)\left(1-\delta_{l_{\rm A},1}\right)
\sum_{l_{\rm B},m_{\rm B},j,p,q}\vert Z_{{\rm T}_{\rm A};l_{\rm A},m_{\rm A},L_{\rm I}}(\nu,K;\Psi_{l_{\rm A}+l_{\rm B},m_{\rm A}+m_{\rm B},j,p,q}) \vert\vert M_{l_{\rm B},m_{\rm B}}^{\rm B}\vert\gamma_{l_{\rm B},m_{\rm B}}^{l_{\rm A},m_{\rm A}}}\nonumber\\
& &\times\frac{1}{a_{\rm B}^{l_{\rm A}+l_{\rm B}+1}}\kappa_{l_{\rm A}+l_{\rm B},j}d_{j,m_{\rm A}+m_{\rm B}}^{l_{\rm A}+l_{\rm B}}\left(\varepsilon_{\rm A}\right)F_{l_{\rm A}+l_{\rm B},j,p}\left(I_{\rm B}\right)G_{l_{\rm A}+l_{\rm B},p,q}\left(e_{\rm B}\right)
\exp\left[i\left(\delta_{{\rm T}_{\rm A};l_{\rm A},m_{\rm A},L_{\rm I}}\!\left(\nu,K;\Psi_{l_{\rm A}+l_{\rm B},m_{\rm A}+m_{\rm B},j,p,q}\right)-\Psi_{l_{\rm A}+l_{\rm B},m_{\rm A}+m_{\rm B},j,p,q}-\delta M_{l_{\rm B},m_{\rm B}}^{\rm B}\right)\right],\nonumber\\
\label{MTA1}
\end{eqnarray}
\begin{eqnarray}
M^{{\rm T}_{\rm A};\rm I\!I}_{l_{\rm A},m_{\rm A}}&=&-\frac{1}{M_{\rm A}}\frac{4\pi}{3}k_{l_{\rm A}}^{\rm A}R_{\rm A}^{2l_{\rm A}+1}\delta_{l_{\rm A},1}\sum_{l^{'}_{\rm A},m^{'}_{\rm A},l_{\rm B},m_{\rm B},r,s,u}\left(-1\right)^{l^{'}_{\rm A}+1}\left(2l^{'}_{\rm A}+2l_{\rm B}+1\right)\vert Z_{{\rm T}_{\rm A};l_{\rm A},m_{\rm A},L_{\rm I\!I}}(\nu,K;\Psi_{l^{'}_{\rm A}+l_{\rm B}+1,m^{'}_{\rm A}+m_{\rm B}+m_{\rm A},r,s,u})\vert\gamma_{l_{\rm B},m_{\rm B}}^{l^{'}_{\rm A},m^{'}_{\rm A}}|M_{l^{'}_{\rm A},m^{'}_{\rm A}}^{\rm A}| |M_{l_{\rm B},m_{\rm B}}^{\rm B}|\nonumber\\
& &\times\gamma^{1,m_{\rm A}}_{l^{'}_{\rm A}+l_{\rm B},m^{'}_{\rm A}+m_{\rm B}}\frac{1}{a_{\rm B}^{l^{'}_{\rm A}+l_{\rm B}+2}}\kappa_{l^{'}_{\rm A}+l_{\rm B}+1,r}
d_{r,m^{'}_{\rm A}+m_{\rm B}+m_{\rm A}}^{l^{'}_{\rm A}+l_{\rm B}+1}\left(\varepsilon_{\rm A}\right)F_{l^{'}_{\rm A}+l_{\rm B}+1,r,s}\left(I_{\rm B}\right) G_{l^{'}_{\rm A}+l_{\rm B}+1,s,u}\left(e_{\rm B}\right)\nonumber\\
& &\times\exp\left[i\left(\delta_{{\rm T}_{\rm A};l_{\rm A},m_{\rm A},L_{\rm I\!I}}\left(\nu,K;\Psi_{l^{'}_{\rm A}+l_{\rm B}+1,m^{'}_{\rm A}+m_{\rm B}+m_{\rm A},r,s,u}\right)-\Psi_{l^{'}_{\rm A}+l_{\rm B}+1,m^{'}_{\rm A}+m_{\rm B}+m_{\rm A},r,s,u}
-\delta M_{l^{'}_{\rm A},m^{'}_{\rm A}}^{\rm A}-\delta M_{l_{\rm B},m_{\rm B}}^{\rm B}\right)\right],\nonumber\\
\label{MTA2}
\end{eqnarray}
{where $M_{l_{\rm B},m_{\rm B}}^{\rm B}=M_{l_{\rm B},m_{\rm B}}^{\rm S_{\rm B}}+M_{l_{\rm B},m_{\rm B}}^{\rm T_{\rm B}}$\footnote{The tidal multipole moments of B due to A can be derived using the same methodology and substituting A to B for the perturber and vice-versa.} 
and $M_{l_{\rm A},m_{\rm A}}^{\rm A}=M_{l_{\rm A},m_{\rm A}}^{\rm S_{\rm A}}+M_{l_{\rm A},m_{\rm A}}^{\rm T_{\rm A}}$.

As for $V_{\rm T}^{\rm A}$, ${\widetilde V}_{\rm T}^{\rm A}$ can be splitted into two components. The first one  ${\widetilde V}_{{\rm T};1}^{\rm A}\left(\vec r,\vec r_{\rm AB}\right)$ is stationary. It corresponds to the permanent component $V_{{\rm T};1}^{\rm A}$ for which the tidal frequency ($\sigma$) vanishes. The second component ${\widetilde V}_{{\rm T};2}^{\rm A}\left(t,\vec r,\vec r_{\rm AB}\right)$ is the time-dependent one that corresponds to $V_{{\rm T};2}^{\rm A}$ for which $\sigma\ne0$.}\\

Finally the external potential of A is thus written in its more compact and general form for $\vert r\vert \ge R_{\rm A}$:
\begin{equation}
V_{\rm ext}^{\rm A}\left(t,\vec r,\vec r_{\rm AB}\right)=G\sum_{l_{\rm A}=0}^{\infty}\sum_{m_{\rm A}=-l_{\rm A}}^{l_{\rm A}}M_{l_{\rm A},m_{\rm A}}^{\rm A}\left(t,\vec r_{\rm AB}\right)\frac{Y_{l_{\rm A},m_{\rm A}}\left(\theta,\varphi\right)}{r^{l_{\rm A}+1}}
\end{equation}
where
\begin{equation}
M^{\rm A}_{l_{\rm A},m_{\rm A}}=M_{l_{\rm A},m_{\rm A}}^{{\rm S}_{\rm A}}+M_{l_{\rm A},m_{\rm A}}^{{\rm T}_{\rm A}}=M_{l_{\rm A},m_{\rm A}}^{{\rm S}_{\rm A}}+M^{{\rm T}_{\rm A};\rm I}_{l_{\rm A},m_{\rm A}}+M^{{\rm T}_{\rm A};\rm I\!I}_{l_{\rm A},m_{\rm A}}.
\label{MA}
\end{equation}

\subsection{Disturbing function}

\begin{figure}[h!]
\centering
\resizebox{12cm}{!}{\includegraphics{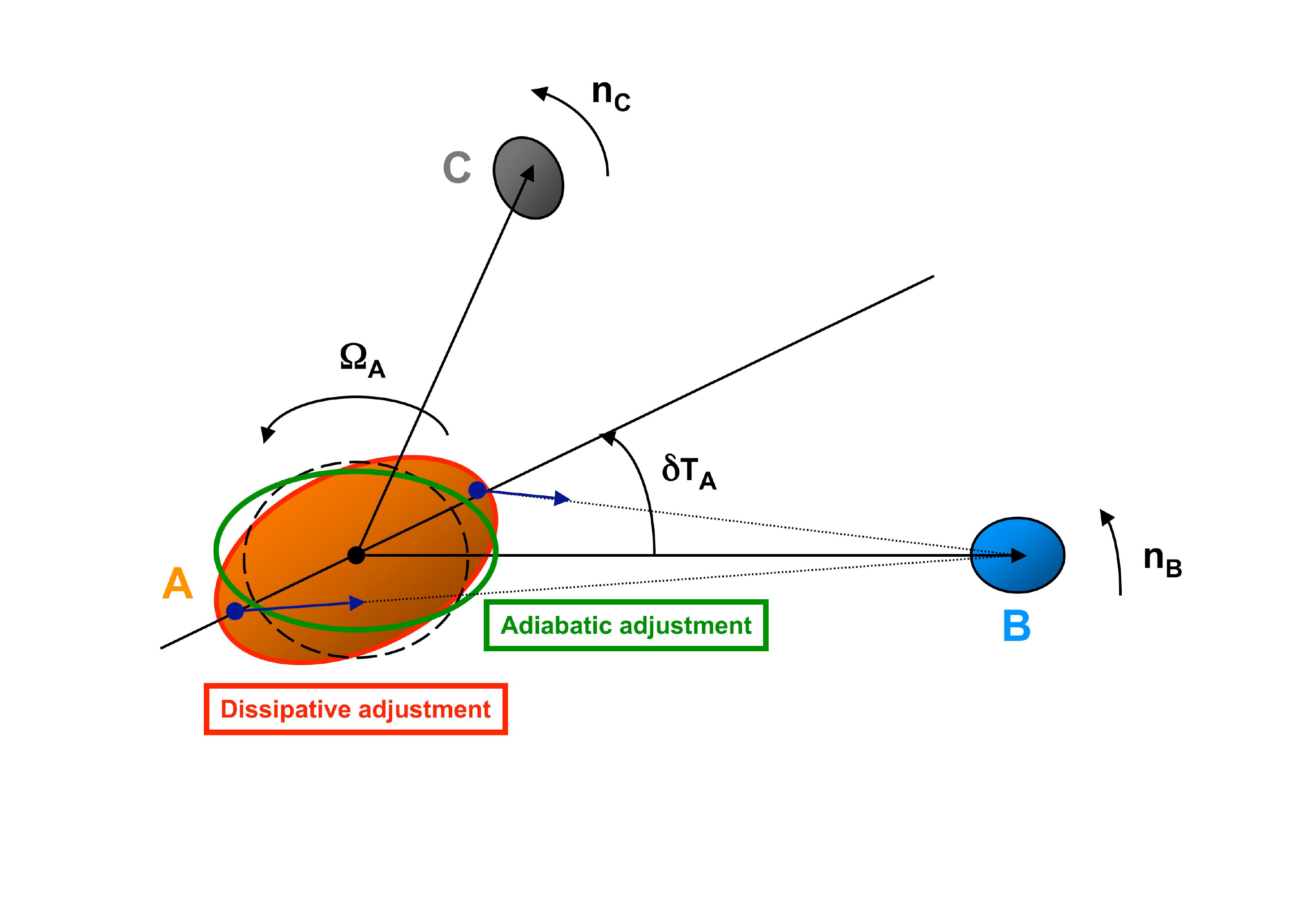}}
\caption{Classical tidal dynamical system. The extended body B is tidally disturbing the extended body A which adjusts itself with a phase lag $\delta_{{\rm T}_{\rm A}}$ due to its internal friction processes. The dynamics of a third body C (different from B or not) is then studied. $\Omega_{\rm A}$, $n_{\rm B}$, $n_{\rm C}$ are respectively the spin frequency of A, and the respective mean motions of B and C.}
\label{fMaree}
\end{figure}

The goal of this section is to derive the disturbing function, ${\mathcal R}_{A-C}$, due to a tidally perturbed body A, acting on a body C of which dynamics is studied and which can be different from the perturber body B (see Fig. (\ref{fMaree})).\\

First, the disturbing function is related to the mutual gravitational interaction potential (cf Tisserand, 1889-1891; Correia 2001) through:
\begin{equation}
{\mathcal R}_{{\rm A}-{\rm C}}\left(t,\vec r_{\rm AC}\right)=-\frac{1}{M_{\rm C}}V_{\rm A-\rm C};
\end{equation}
the sign being due to the potentials convention adopted here.\\

Using the definition of $V_{\rm A-\rm C}$ given in Eq. (\ref{Uint}), we deduce the explicit spectral expansion of ${\mathcal R}_{{\rm A}-{\rm C}}$ in the spherical harmonics:
\begin{equation}
{\mathcal R}_{{\rm A}-{\rm C}}=-\frac{G}{M_{\rm C}}\sum_{l_{\rm A}=0}^{\infty}\sum_{m_{\rm A}=-l_{\rm A}}^{l_{\rm A}}\sum_{l_{\rm C}=0}^{\infty}\sum_{m_{\rm C}=-l_{\rm C}}^{l_{\rm C}}\left\{M_{l_{\rm A},m_{\rm A}}^{\rm A}M_{l_{\rm C},m_{\rm C}}^{\rm C}
\left(-1\right)^{l_{\rm A}}\gamma_{l_{\rm C},m_{\rm C}}^{l_{\rm A},m_{\rm A}}\frac{Y_{l_{\rm A}+l_{\rm C},m_{\rm A}+m_{\rm C}}\left(\theta_{\rm AC},\varphi_{\rm AC}\right)}{r_{\rm AC}^{l_{\rm A}+l_{\rm C}+1}}\right\}.
\end{equation}
The $M^{\rm A}_{l_{\rm A},m_{\rm A}}$, $M^{\rm C}_{l_{\rm C},m_{\rm C}}$ are respectively the mass multipole moments of the body A and of the body C, while $r_{\rm AC}$, $\theta_{\rm AC}$ and $\varphi_{\rm AC}$ are the spherical coordinates of the center of mass of the body C in the A-frame (cf. Fig (\ref{StructureTide})). Then, using the Kaula's transformation given in Eqs. (\ref{Kaula1}), (\ref{Kaula2}) and (\ref{Kaula3}), ${\mathcal R}_{\rm A-\rm C}$ is expressed as a function of the obliquity, $\varepsilon_{\rm A}$, and of the Keplerian orbital elements of C: $a_{\rm C}$, $e_{\rm C}$ and $I_{\rm C}$:
\begin{eqnarray}
{\mathcal R}_{{\rm A}-{\rm C}}&=&-\frac{G}{M_{\rm C}}\sum_{l_{\rm A}=0}^{\infty}\sum_{m_{\rm A}=-l_{\rm A}}^{l_{\rm A}}\sum_{l_{\rm C}=0}^{\infty}\sum_{m_{\rm C}=-l_{\rm C}}^{l_{\rm C}}\left\{M_{l_{\rm A},m_{\rm A}}^{\rm A}M_{l_{\rm C},m_{\rm C}}^{\rm C}\left(-1\right)^{l_{\rm A}}\gamma_{l_{\rm C},m_{\rm C}}^{l_{\rm A},m_{\rm A}}\right.\nonumber\\
& &\times{\left.\frac{1}{a_{\rm C}^{l_{\rm A}+l_{\rm C}+1}}\sum_{v=-\left(l_{\rm A}+l_{\rm C}\right)}^{\left(l_{\rm A}+l_{\rm C}\right)}\sum_{w=0}^{l_{\rm A}+l_{\rm C}}\sum_{b}\kappa_{l_{\rm A}+l_{\rm C},v}
d_{v,m_{\rm A}+m_{\rm C}}^{l_{\rm A}+l_{\rm C}}\left(\varepsilon_{\rm A}\right)F_{l_{\rm A}+l_{\rm C},v,w}\left(I_{\rm C}\right)G_{l_{\rm A}+l_{\rm C},w,b}\left(e_{\rm C}\right)\exp\left[i\Psi_{l_{\rm A}+l_{\rm C},m_{\rm A}+m_{\rm C},v,w,b}\right]\right\}}.
\label{RAC0}
\end{eqnarray}

\par Here, three types of gravitational interaction are treated in our formalism (see also Eq. (\ref{MA})). To describe them, one has first to consider the two causes of the multipolar behaviour of the gravitational potential of a body. The first is due to its internal structure and dynamics. In the case of a solid body, it is due to its proper asymmetry while in the case of a fluid mass, the internal dynamical processes such as rotation or magnetic field will break the ideal spherical hydrostatic symmetry of the body. The second is the deformation of the body due to its response to the tidal perturbation exerted by the perturber(s). In the case studied here, it is the response of the body A to the perturbation exerted by B computed in the previous section. Therefore, we split here the $k$-indexed mass multipole moments of each body as in Eq. (\ref{GME}):    
\begin{equation}
M_{l_{k},m_{k}}^{k}=M_{l_k,m_k}^{{\rm S}_k}+M_{l_k,m_k}^{{\rm T}_k},
\label{MultipoleNature}
\end{equation}
where $M_{l_k,m_k}^{{\rm S}_k}$ is the self-structural contribution of the body while $M_{l_k,m_k}^{{\rm T}_k}$ is the tidal one.\\

The three type of gravitational interaction are thus identified. The first is the interaction between the structural mass multipole moments of each body, $M_{l_k,m_k}^{{\rm S}_k}M_{l_{k^{'}},m_{k^{'}}}^{{\rm S}_{{k}^{'}}}$ with $k \ne k^{'}$; one should note that $\left\{l_{\rm A}=0,m_{\rm A}=0\right\}$-$\left\{l_{\rm C}=0,m_{\rm C}=0\right\}$ is the classical interaction between $M_{\rm A}$ and $M_{\rm C}$, $M_{\rm C}$ being the mass of C. The second corresponds to the mixed interaction between the structural and the tidal mass multipole moments, $M_{l_k,m_k}^{{\rm S}_k}M_{l_{k^{'}},m_{k^{'}}}^{{\rm T}_{k^{'}}}$. The third is the interaction between the tidal mass multipole moments of each body, $M_{l_k,m_k}^{{\rm T}_k}M_{l_{k^{'}},m_{k^{'}}}^{{\rm T}_{k^{'}}}$. Therefore, the disturbing function could be splitted into three terms:
\begin{equation}
{\mathcal R}_{{\rm A}-{\rm C}}={\mathcal R}_{{\rm A}-{\rm C};{\rm S}-{\rm S}}\left(t,{\vec r}_{\rm AC}\right)+{\mathcal R}_{{\rm A}-{\rm C};{\rm S}-{\rm T}}\left(t,{\vec r}_{\rm AC}\right)+{\mathcal R}_{{\rm A}-{\rm C};{\rm T}-{\rm T}}\left(t,{\vec r}_{\rm AC}\right),
\end{equation}
where ${\mathcal R}_{{\rm A}-{\rm C};{\rm S}-{\rm S}}$ is the disturbing function associated to the structure-structure interaction, ${\mathcal R}_{{\rm A}-{\rm C};{\rm T}-{\rm S}}$ is associated to the tide-structure interaction and ${\mathcal R}_{{\rm A}-{\rm C};{\rm T}-{\rm T}}$, is associated to the tide-tide interaction.\\

Inserting Eq. (\ref{MultipoleNature}) into Eq. (\ref{RAC0}), the respective Fourier expansions of ${\mathcal R}_{\rm A-\rm C;\rm S-\rm S}$, ${\mathcal R}_{\rm A-\rm C;\rm S-\rm T}$ and ${\mathcal R}_{\rm A-\rm C;\rm T-\rm T}$ are derived:
\begin{eqnarray}
{\mathcal R}_{{\rm A}-{\rm C};{\rm S}-{\rm S}}&=&-\frac{G}{M_{\rm C}}\sum_{l_{\rm A}=0}^{\infty}\sum_{m_{\rm A}=-l_{\rm A}}^{l_{\rm A}}\sum_{l_{\rm C}=0}^{\infty}\sum_{m_{\rm C}=-l_{\rm C}}^{l_{\rm C}}\left\{M_{l_{\rm A},m_{\rm A}}^{{\rm S}_{\rm A}}M_{l_{\rm C},m_{\rm C}}^{{\rm S}_{\rm C}}\left(-1\right)^{l_{\rm A}}\gamma_{l_{\rm C},m_{\rm C}}^{l_{\rm A},m_{\rm A}}\right.\nonumber\\
& &\times{\left.\frac{1}{a_{\rm C}^{l_{\rm A}+l_{\rm C}+1}}\sum_{v,w,b}\kappa_{l_{\rm A}+l_{\rm C},v}d_{v,m_{\rm A}+m_{\rm C}}^{l_{\rm A}+l_{\rm C}}\left(\varepsilon_{\rm A}\right)
F_{l_{\rm A}+l_{\rm C},v,w}\left(I_{\rm C}\right)G_{l_{\rm A}+l_{\rm C},w,b}\left(e_{\rm C}\right)
\exp\left[i\Psi_{l_{\rm A}+l_{\rm C},m_{\rm A}+m_{\rm C},v,w,b}\right]\right\}},
\end{eqnarray}
\begin{eqnarray}
{\mathcal R}_{{\rm A}-{\rm C};{\rm T}-{\rm S}}&=&-\frac{G}{M_{\rm C}}\sum_{l_{\rm A}=0}^{\infty}\sum_{m_{\rm A}=-l_{\rm A}}^{l_{\rm A}}\sum_{l_{\rm C}=0}^{\infty}\sum_{m_{\rm C}=-l_{\rm C}}^{l_{\rm C}}\left\{(M_{l_{\rm A},m_{\rm A}}^{{\rm S}_{\rm A}}M_{l_{\rm C},m_{\rm C}}^{{\rm T}_{\rm C}}+M_{l_{\rm A},m_{\rm A}}^{{\rm T}_{\rm A}}M_{l_{\rm C},m_{\rm C}}^{{\rm S}_{\rm C}})\left(-1\right)^{l_{\rm A}}\gamma_{l_{\rm C},m_{\rm C}}^{l_{\rm A},m_{\rm A}}\right.\nonumber\\
& &\times{\left.\frac{1}{a_{\rm C}^{l_{\rm A}+l_{\rm C}+1}}\sum_{v,w,b}\kappa_{l_{\rm A}+l_{\rm C},v}
d_{v,m_{\rm A}+m_{\rm C}}^{l_{\rm A}+l_{\rm C}}\left(\varepsilon_{\rm A}\right) F_{l_{\rm A}+l_{\rm C},v,w}\left(I_{\rm C}\right)G_{l_{\rm A}+l_{\rm C},w,b}\left(e_{\rm C}\right)
\exp\left[i\Psi_{l_{\rm A}+l_{\rm C},m_{\rm A}+m_{\rm C},v,w,b}\right]\right\}}
\end{eqnarray}
and
\begin{eqnarray}
{\mathcal R}_{{\rm A}-{\rm C};{\rm T}-{\rm T}}&=&-\frac{G}{M_{\rm C}}\sum_{l_{\rm A}=0}^{\infty}\sum_{m_{\rm A}=-l_{\rm A}}^{l_{\rm A}}\sum_{l_{\rm C}=0}^{\infty}\sum_{m_{\rm C}=-l_{\rm C}}^{l_{\rm C}}\left\{M_{l_{\rm A},m_{\rm A}}^{{\rm T}_{\rm A}}M_{l_{\rm C},m_{\rm C}}^{{\rm T}_{\rm C}}
\left(-1\right)^{l_{\rm A}}\gamma_{l_{\rm C},m_{\rm C}}^{l_{\rm A},m_{\rm A}}\right.\nonumber\\
&&\times{\left.\frac{1}{a_{\rm C}^{l_{\rm A}+l_{\rm C}+1}}\sum_{v,w,b}\kappa_{l_{\rm A}+l_{\rm C},v}d_{v,m_{\rm A}+m_{\rm C}}^{l_{\rm A}+l_{\rm C}}\left(\varepsilon_{\rm A}\right)
F_{l_{\rm A}+l_{\rm C},v,w}\left(I_{\rm C}\right)G_{l_{\rm A}+l_{\rm C},w,b}\left(e_{\rm C}\right)
\exp\left[i\Psi_{l_{\rm A}+l_{\rm C},m_{\rm A}+m_{\rm C},v,w,b}\right]\right\}}.
\end{eqnarray}

This classification of the three different types of interaction allows  to explicitly generalize the classical case where the only considered extended body is the tidally perturbed one, A, while B and C are considered as ponctual masses. In this case, the interaction are restricted to the classical gravitational interaction between $M^{{\rm S}_{\rm A}}_{l_{\rm A},m_{\rm A}}$, $M^{{\rm T}_{\rm A}}_{l_{\rm A},m_{\rm A}}$, $M_{\rm B}$, the mass of B, and $M_{\rm C}$.\\

{By now, to lighten the equations, the tidal multipole moments of C are ignored. In a practical case, they have to be derived using eqs. (\ref{MTA1}-\ref{MTA2}) and taken into account.} 

The disturbing function ${\mathcal R}_{\rm A-\rm C}$ is thus reduced to the two first interactions: the respective structural moments of body A and of body C and the structural moments of body C with the tidal moments of body A. The Fourier expansion of the disturbing function is thus given by:
\begin{equation}
{\mathcal R}_{{\rm A}-{\rm C}}={\mathcal R}_{{\rm A}-{\rm C};{\rm S}-{\rm S}}+{\mathcal R}_{{\rm A}-{\rm C};{\rm T}-{\rm S}}=\sum_{l_{\rm A},m_{\rm A},l_{\rm C},m_{\rm C},v,w,b}{\mathcal R}_{l_{\rm A},m_{\rm A},l_{\rm C},m_{\rm C},v,w,b}\left(t,{\vec r}_{\rm AC}\right),
\end{equation}
where
\begin{equation}
{\mathcal R}_{l_{\rm A},m_{\rm A},l_{\rm C},m_{\rm C},v,w,b}={\mathcal R}_{{\rm S}-{\rm S};l_{\rm A},m_{\rm A},l_{\rm C},m_{\rm C},v,w,b}\left(t,{\vec r}_{\rm AC}\right)+{\mathcal R}_{{\rm T}-{\rm S};l_{\rm A},m_{\rm A},l_{\rm C},m_{\rm C},v,w,b}\left(t,{\vec r}_{\rm AC}\right)
\label{RAC1}
\end{equation}
with
\begin{eqnarray}
{\mathcal R}_{{\rm S}-{\rm S};l_{\rm A},m_{\rm A},l_{\rm C},m_{\rm C},v,w,b}&=&-\frac{G}{M_{\rm C}}M_{l_{\rm A},m_{\rm A}}^{{\rm S}_{\rm A}}M_{l_{\rm C},m_{\rm C}}^{{\rm S}_{\rm C}}\left(-1\right)^{l_{\rm A}}\gamma_{l_{\rm C},m_{\rm C}}^{l_{\rm A},m_{\rm A}}\nonumber\\ 
& &\times\frac{1}{a_{\rm C}^{l_{\rm A}+l_{\rm C}+1}} \sum_{v,w,b}\kappa_{l_{\rm A}+l_{\rm C},v}d_{v,m_{\rm A}+m_{\rm C}}^{l_{\rm A}+l_{\rm C}}\left(\varepsilon_{\rm A}\right)F_{l_{\rm A}+l_{\rm C},v,w}\left(I_{\rm C}\right)
G_{l_{\rm A}+l_{\rm C},w,b}\left(e_{\rm C}\right)\exp\left[i\Psi_{l_{\rm A}+l_{\rm C},m_{\rm A}+m_{\rm C},v,w,b}\right]
\label{RAC2}
\end{eqnarray}
and
\begin{eqnarray}
{\mathcal R}_{{\rm T}-{\rm S};l_{\rm A},m_{\rm A},l_{\rm C},m_{\rm C},v,w,b}&=&-\frac{G}{M_{\rm C}}M_{l_{\rm A},m_{\rm A}}^{{\rm T}_{\rm A}}M_{l_{\rm C},m_{\rm C}}^{{\rm S}_{\rm C}}\left(-1\right)^{l_{\rm A}}\gamma_{l_{\rm C},m_{\rm C}}^{l_{\rm A},m_{\rm A}}\nonumber\\ 
& &\times \frac{1}{a_{\rm C}^{l_{\rm A}+l_{\rm C}+1}}\sum_{v,w,b}\kappa_{l_{\rm A}+l_{\rm C},v}d_{v,m_{\rm A}+m_{\rm C}}^{l_{\rm A}+l_{\rm C}}\left(\varepsilon_{\rm A}\right)F_{l_{\rm A}+l_{\rm C},v,w}\left(I_{\rm C}\right)
G_{l_{\rm A}+l_{\rm C},w,b}\left(e_{\rm C}\right)\exp\left[i\Psi_{l_{\rm A}+l_{\rm C},m_{\rm A}+m_{\rm C},v,w,b}\right].
\label{RAC3}
\end{eqnarray}
Using Eq. (\ref{MTA}), the Tide-Structure interaction disturbing function is expanded as follow:
\begin{eqnarray}
{\mathcal R}_{{\rm T}-{\rm S};l_{\rm A},m_{\rm A},l_{\rm C},m_{\rm C},v,w,b}=\sum_{l_{\rm B},m_{\rm B},j,p,q}{\mathcal R}_{{\rm I}; L_{\rm I}}\left(t,{\vec r}_{\rm AC}\right)+\sum_{l^{'}_{\rm A},m^{'}_{\rm A},l_{\rm B},m_{\rm B},r,s,u}{\mathcal R}_{{\rm I\!I}; L_{\rm I\!I}}\left(t,{\vec r}_{\rm AC}\right),
\label{RAC4}
\end{eqnarray}
where ${\mathcal R}_{{\rm I}; L_{\rm I}}$ and ${\mathcal R}_{{\rm I\!I}; L_{\rm I\!I}}$ respectively correspond to the $M^{{\rm T}_{\rm A};\rm I}_{l_{\rm A},m_{\rm A}}$ and $M^{{\rm T}_{\rm A};\rm I\!I}_{l_{\rm A},m_{\rm A}}$ contribution to the external gravitational potential of the tidally perturbed body A, $V_{\rm T}^{\rm A}$. Inserting the explicit expansion for $M^{{\rm T}_{\rm A};\rm I}_{l_{\rm A},m_{\rm A}}$ and $M^{{\rm T}_{\rm A};\rm I\!I}_{l_{\rm A},m_{\rm A}}$ given in Eqs. (\ref{MTA1}) and (\ref{MTA2}) into (\ref{RAC3}), we get 
\begin{eqnarray}
{\mathcal R}_{{\rm I}; L_{\rm I}}&=&Z_{{\rm T}_{\rm A};l_{\rm A},m_{\rm A},L_{\rm I}}(\nu,K;\Psi_{l_{\rm A}+l_{\rm B},m_{\rm A}+m_{\rm B},j,p,q}) {\mathcal R}_{{\rm I}; L_{\rm I}}^{\rm Ad}\nonumber\\
&=&-\frac{G}{M_{\rm C}}\frac{4\pi}{2l_{\rm A}+1}k_{l_{\rm A}}^{\rm A}R_{\rm A}^{2l_{\rm A}+1}\left(1-\delta_{l_{\rm A},0}\right)\left(1-\delta_{l_{\rm A},1}\right)\vert Z_{{\rm T}_{\rm A};l_{\rm A},m_{\rm A},L_{\rm I}}(\nu,K;\Psi_{l_{\rm A}+l_{\rm B},m_{\rm A}+m_{\rm B},j,p,q}) \vert \gamma_{l_{\rm B},m_{\rm B}}^{l_{\rm A},m_{\rm A}}|M^{\rm B}_{l_{\rm B},m_{\rm B}}||M^{{\rm S}_{\rm C}}_{l_{\rm C},m_{\rm C}}|\gamma_{l_{\rm C},m_{\rm C}}^{l_{\rm A},m_{\rm A}}\nonumber\\
& &\times \frac{1}{a_{B}^{l_{\rm A}+l_{\rm B}+1}}\frac{1}{a_{\rm C}^{l_{\rm A}+l_{\rm C}+1}}\kappa_{l_{\rm A}+l_{\rm B},j}\kappa_{l_{\rm A}+l_{\rm C},v} d_{j,m_{\rm A}+m_{\rm B}}^{l_{\rm A}+l_{\rm B}}\left(\varepsilon_{\rm A}\right)d_{v,m_{\rm A}+m_{\rm C}}^{l_{\rm A}+l_{\rm C}}\left(\varepsilon_{\rm A}\right)F_{l_{\rm A}+l_{\rm B},j,p}\left(I_{\rm B}\right)F_{l_{\rm A}+l_{\rm C},v,w}\left(I_{\rm C}\right)
G_{l_{\rm A}+l_{\rm B},p,q}\left(e_{\rm B}\right)G_{l_{\rm A}+l_{\rm C},w,b}\left(e_{\rm C}\right)\nonumber\\
& &\times\exp\left[i\left(\Psi_{l_{\rm A}+l_{\rm C},m_{\rm A}+m_{\rm C},v,w,b}-\Psi_{l_{\rm A}+l_{\rm B},m_{\rm A}+m_{\rm B},j,p,q}+\delta M_{l_{\rm C},m_{\rm C}}^{{\rm S}_{\rm C}}-\delta M_{l_{\rm B},m_{\rm B}}^{\rm B}+\delta_{{\rm T}_{\rm A};l_{\rm A},m_{\rm A},L_{\rm I}}\left(\nu,K;\Psi_{l_{\rm A}+l_{\rm B},m_{\rm A}+m_{\rm B},j,p,q}\right)\right)\right],
\label{RAC5}
\end{eqnarray}
where $L_{\rm I}=\left\{l_{\rm B},m_{\rm B},j,p,q\right\}$, and for ${\mathcal R}_{{\rm I\!I}; L_{\rm I\!I}}$:
\begin{eqnarray}
{\mathcal R}_{{\rm I\!I}; L_{\rm I\!I}}&=&Z_{{\rm T}_{\rm A}; l_{\rm A},m_{\rm A},L_{\rm I\!I}}(\nu,K;\Psi_{l^{'}_{\rm A}+l_{\rm B}+1,m^{'}_{\rm A}+m_{\rm B}+m_{\rm A},r,s,u}){\mathcal R}_{{\rm I\!I}; L_{\rm I\!I}}^{\rm Ad}\nonumber\\ 
&=&\frac{G}{M_{\rm C}}\frac{1}{M_{\rm A}}\frac{4\pi}{3}k^{\rm A}_{l_{\rm A}}R_{\rm A}^{2l_{\rm A}+1}\delta_{l_{\rm A},1}\left(-1\right)^{l^{'}_{\rm A}+1+l_{\rm A}}
\left(2l^{'}_{\rm A}+2l_{\rm B}+1\right)\vert Z_{{\rm T}_{\rm A}; l_{\rm A},m_{\rm A},L_{\rm I\!I}}(\nu,K;\Psi_{l^{'}_{\rm A}+l_{\rm B}+1,m^{'}_{\rm A}+m_{\rm B}+m_{\rm A},r,s,u}) \vert\gamma_{l_{\rm B},m_{\rm B}}^{l^{'}_{\rm A},m^{'}_{\rm A}}|M^{\rm A}_{l^{'}_{\rm A},m^{'}_{\rm A}}||M^{\rm B}_{l_{\rm B},m_{\rm B}}|\nonumber\\
& &\times\gamma_{l^{'}_{\rm A}+l_{\rm B},m^{'}_{\rm A}+m_{\rm B}}^{1,m_{\rm A}}|M^{{\rm S}_{\rm C}}_{l_{\rm C},m_{\rm C}}|
\gamma_{l_{\rm C},m_{\rm C}}^{l_{\rm A},m_{\rm A}} \frac{1}{a_{\rm B}^{l^{'}_{\rm A}+l_{\rm B}+2}}\frac{1}{a_{\rm C}^{l_{\rm A}+l_{\rm C}+1}}\kappa_{l^{'}_{\rm A}+l_{\rm B}+1,r}\kappa_{l_{\rm A}+l_{\rm C},v}d_{r,m^{'}_{\rm A}+m_{\rm B}+m_{\rm A}}^{l^{'}_{\rm A}+l_{\rm B}+1}\left(\varepsilon_{\rm A}\right)d_{v,m_{\rm A}+m_{\rm C}}^{l_{\rm A}+l_{\rm C}}\left(\varepsilon_{\rm A}\right)\nonumber\\
& &\times F_{l^{'}_{\rm A}+l_{\rm B}+1,r,s}\left(I_{\rm B}\right)F_{l_{\rm A}+l_{\rm C},v,w}\left(I_{\rm C}\right)G_{l^{'}_{\rm A}+l_{\rm B}+1,s,u}\left(e_{\rm B}\right)G_{l_{\rm A}+l_{\rm C},w,b}\left(e_{\rm C}\right)\nonumber\\
& &\times \exp\left[i\left(\Psi_{l_{\rm A}+l_{\rm C},m_{\rm A}+m_{\rm C},v,w,b}\!-\!\Psi_{l^{'}_{\rm A}+l_{\rm B}+1,m^{'}_{\rm A}+m_{\rm B}+m_{\rm A},r,s,u}\!+\!\delta M^{{\rm S}_{\rm C}}_{l_{\rm C},m_{\rm C}}\!-\!\delta M_{l^{'}_{\rm A},m^{'}_{\rm A}}^{\rm A}\!-\!\delta M_{l_{\rm B},m_{\rm B}}^{\rm B}\!+\!\delta_{{\rm T}_{\rm A};l_{\rm A},m_{\rm A},L_{\rm I\!I}}\left(\nu,K;\Psi_{l^{'}_{\rm A}+l_{\rm B}+1,m^{'}_{\rm A}+m_{\rm B}+m_{\rm A},r,s,u}\right)\right)\right],\nonumber\\
\label{RAC6}
\end{eqnarray}
where $L_{\rm I\!I}=\left\{l^{'}_{\rm A},m^{'}_{\rm A},l_{\rm B},m_{\rm B},r,s,u\right\}$. ${\mathcal R}_{{\rm I}; L_{\rm I}}^{\rm Ad}$ and ${\mathcal R}_{{\rm I\!I}; L_{\rm I\!I}}^{\rm Ad}$ correspond to the adiabatic response of A. {Finally, as in Eq. (\ref{GME}) $M_{l_{\rm B},m_{\rm B}}^{\rm B}=M_{l_{\rm B},m_{\rm B}}^{\rm S_{\rm B}}+M_{l_{\rm B},m_{\rm B}}^{\rm T_{\rm B}}$ and $M_{l_{\rm A},m_{\rm A}}^{\rm A}=M_{l_{\rm A},m_{\rm A}}^{\rm S_{\rm A}}+M_{l_{\rm A},m_{\rm A}}^{\rm T_{\rm A}}$.}
%In this section, the general disturbing function acting on an extended body C due to a tidally perturbed extended body A has thus been obtained. The next step is to derive the associated equations for the dynamical evolution of this system.
\subsection{Dynamical equations}
The external potential of the tidally perturbed body A by a body B being now well understood and known, our purpose here is to derive the dynamical equations for the evolution of the angular velocity (the angular momentum in term of Andoyer's variables) and the obliquity of A under the action of the gravitational interaction with a body $C$ which could be different from the perturber B and of the Keplerian orbital elements of this third body: $a_{\rm C}$, $e_{\rm C}$ and $I_{\rm C}$. To achieve this, we follow the method adopted by Yoder (1995-1997) and Correia \& Laskar (2003 a,b,c) who used the mutual interaction potential for the variation of the Andoyer's variables and the disturbing function of the orbital elements. {Here, gravitational interactions between B and C are not taken into account.}\\

Begining with the Andoyer's variables (cf. Andoyer, 1926), we respectively get the total angular momentum, $L_{\rm A}=I_{\rm A}\Omega_{\rm A}$, $I_{\rm A}$ being the inertia momentum of A:  
\begin{equation}
I_{\rm A}\frac{{\rm d}\Omega_{\rm A}}{{\rm d}t}=\partial_{\Theta_{\rm A}}V_{\rm A-\rm C},
\end{equation}
and the obliquity, $\varepsilon_{\rm A}$:
\begin{equation}
I_{\rm A}\Omega_{\rm A}\frac{{\rm d}}{{\rm d}t}\cos\varepsilon_{\rm A}=-\partial_{\phi_{\rm A}}V_{\rm A-\rm C}-\cos\varepsilon_{\rm A}\partial_{\Theta_{\rm A}}V_{\rm A-\rm C}.
\end{equation}
The classical equations of orbital evolution are given by the Lagrange's planetary equations (cf Brouwer \& Clemence, 1961):
\begin{equation}
\frac{{\rm d} a_{\rm C}}{{\rm d}t}=\frac{2}{n_{\rm C}a_{\rm C}}\partial_{M_{\rm C}}{\mathcal R_{\rm A-\rm C}},
\end{equation}
\begin{equation}
\frac{{\rm d}e_{\rm C}}{{\rm d}t}=-\frac{\sqrt{1-e_{\rm C}^2}}{n_{\rm C}a_{\rm C}^2e_{\rm C}}\partial_{\omega_{\rm C}}{\mathcal R_{\rm A-\rm C}}+\frac{1-e_{\rm C}^2}{n_{\rm C}a_{\rm C}^2e_{\rm C}}\partial_{M_{\rm C}}{\mathcal R_{\rm A-\rm C}},
\end{equation}
\begin{equation}
\frac{{\rm d}I_{\rm C}}{{\rm d}t}=-\frac{1}{n_{\rm C}a_{\rm C}^2\sqrt{1-e_{\rm C}^2}\sin I_{\rm C}}\partial_{\Omega_{\rm C}^{*}}{\mathcal R_{\rm A-\rm C}}+\frac{\cos I_{\rm C}}{n_{\rm C} a_{\rm C}^2 \sqrt{1-e_{\rm C}^2} \sin I_{\rm C}}\partial_{\omega_{\rm C}}{\mathcal R_{\rm A-\rm C}}.
\end{equation}
The Fourier expansion of the disturbing function ${\mathcal R}_{{\rm A}-{\rm C}}$ is then introduced:
\begin{equation}
{\mathcal R}_{{\rm A}-{\rm C}}=\sum_{l_{\rm A},m_{\rm A},l_{\rm C},m_{\rm C},v,w,b}{\mathcal R}_{l_{\rm A},m_{\rm A},l_{\rm C},m_{\rm C},v,w,b},
\end{equation}
where, as in Eq. (\ref{RAC0}) , ${\mathcal R}_{l_{\rm A},m_{\rm A},l_{\rm C},m_{\rm C},v,w,b}$ is given by:
\begin{eqnarray}
{\mathcal R}_{l_{\rm A},m_{\rm A},l_{\rm C},m_{\rm C},v,w,b}&=&-\frac{G}{M_{\rm C}} M_{l_{\rm A},m_{\rm A}}^{\rm A}M_{l_{\rm C},m_{\rm C}}^{\rm C}\left(-1\right)^{l_{\rm A}}\gamma_{l_{\rm C},m_{\rm C}}^{l_{\rm A},m_{\rm A}}\nonumber\\
&&\times\frac{1}{a_{\rm C}^{l_{\rm A}+l_{\rm C}+1}}\kappa_{l_{\rm A}+l_{\rm C},v}
d_{v,m_{\rm A}+m_{\rm C}}^{l_{\rm A}+l_{\rm C}}\left(\varepsilon_{\rm A}\right)F_{l_{\rm A}+l_{\rm C},v,w}\left(I_{\rm C}\right)G_{l_{\rm A}+l_{\rm C},w,b}\left(e_{\rm C}\right)\exp\left[i\Psi_{l_{\rm A}+l_{\rm C},m_{\rm A}+m_{\rm C},v,w,b}\right].
\end{eqnarray}
Therefore, the following formal equations are obtained for the dynamical evolution of respectively the angular velocity of A and its obliquity:
\begin{equation}
I_{\rm A}\frac{{\rm d}\Omega_{\rm A}}{{\rm d}t}=M_{\rm C}\sum_{l_{\rm A},m_{\rm A},l_{\rm C},m_{\rm C},v,w,b}{\rm R}_{\rm e}\left\{i\left(m_{\rm A}+m_{\rm C}\right){\mathcal R}_{l_{\rm A},m_{\rm A},l_{\rm C},m_{\rm C},v,w,b}\right\},
\end{equation}
\begin{equation}
I_{\rm A}\Omega_{\rm A}\frac{{\rm d}}{{\rm d}t}\cos\varepsilon_{\rm A}=-M_{\rm C}\sum_{l_{\rm A},m_{\rm A},l_{\rm C},m_{\rm C},v,w,b}{\rm R}_{\rm e}\left\{i\left[v+\left(m_A+m_C\right)\cos\varepsilon_{\rm A}\right]
{\mathcal R}_{l_{\rm A},m_{\rm A},l_{\rm C},m_{\rm C},v,w,b}\right\},
\end{equation}
where ${\rm R}_{e}\left(z\right)$ is the real part of a complex number or function $z$, while those for the keplerian elements are given by: 
\begin{equation}
\frac{1}{a_{\rm C}}\frac{{\rm d}a_{\rm C}}{{\rm d}t}=\frac{2}{n_{\rm C}}\frac{1}{a_{\rm C}^{2}}\sum_{l_{\rm A},m_{\rm A},l_{\rm C},m_{\rm C},v,w,b}{\rm R}_{\rm e}\left\{i\left(l_{\rm A}+l_{\rm C}-2w+b\right)
{\mathcal R}_{l_{\rm A},m_{\rm A},l_{\rm C},m_{\rm C},v,w,b}\right\},
\end{equation}
\begin{equation}
\frac{1}{e_{\rm C}}\frac{{\rm d}e_{\rm C}}{{\rm d}t}=\frac{1}{n_{\rm C}}\frac{1-e_{\rm C}^{2}}{e_{\rm C}^{2}}\frac{1}{a_{\rm C}^{2}}
\sum_{l_{\rm A},m_{\rm A},l_{\rm C},m_{\rm C},v,w,b}{\rm R}_{\rm e}\left\{i\left[\left(l_{\rm A}+l_{\rm C}-2w\right)\left(1-\frac{1}{\sqrt{1-e_{\rm C}^{2}}}\right)+b\right]
{\mathcal R}_{l_{\rm A},m_{\rm A},l_{\rm C},m_{\rm C},v,w,b}\right\}
\end{equation}
and
\begin{equation}
\frac{{\rm d}}{{\rm d}t}\cos I_{\rm C}=\frac{1}{n_{\rm C}}\frac{1}{\sqrt{1-e_{\rm C}^{2}}}\frac{1}{a_{\rm C}^{2}}\sum_{l_{\rm A},m_{\rm A},l_{\rm C},m_{\rm C},v,w,b}{\rm R}_{\rm e}\left\{i\left[v-\left(l_{\rm A}+l_{\rm C}-2w\right)\cos I_{\rm C}\right]
{\mathcal R}_{l_{\rm A},m_{\rm A},l_{\rm C},m_{\rm C},v,w,b}\right\}.
\end{equation}
If {the tidal multipole moments of C are ignored}, the following system is obtained using Eqs. (\ref{RAC1}) and (\ref{RAC4}) due to the dependance of ${\mathcal R}_{{\rm I};L_{\rm I}}$ and ${\mathcal R}_{{\rm I\!I};L_{\rm I\!I}}$ on $\Theta_{\rm A}$ and $\phi_{\rm A}$:
\begin{eqnarray}
\lefteqn{I_{\rm A}\frac{{\rm d}{\Omega}_{\rm A}}{{\rm d}t}=M_{\rm C}\sum_{l_{\rm A},m_{\rm A},l_{\rm C},m_{\rm C},v,w,b}{\rm R}_{\rm e}\left\{i\left(m_{\rm A}+m_{\rm C}\right){\mathcal R}_{{\rm S}-{\rm S};l_{\rm A},m_{\rm A},l_{\rm C},m_{\rm C},v,w,b}+i\sum_{l_{\rm B},m_{\rm B},j,p,q}\left(m_{\rm C}-m_{\rm B}\right){\mathcal R}_{{\rm I}; L_{\rm I}}\right.}\nonumber\\
& &{\left.+i\sum_{l^{'}_{\rm A},m^{'}_{\rm A},l_{\rm B},m_{\rm B},r,s,u}\left(m_{\rm C}-\left(m^{'}_{\rm A}+m_{\rm B}\right)\right){\mathcal R}_{{\rm I\!I}; L_{\rm I\!I}}-\sum_{l_{\rm B},m_{\rm B},j,p,q}\partial_{\Theta_{\rm A}}\left[Z_{{\rm T}_{\rm A};l_{\rm A},m_{\rm A},L_{\rm I}}\right]{\mathcal R}_{{\rm I}; L_{\rm I}}^{\rm Ad}
-\sum_{l^{'}_{\rm A},m^{'}_{\rm A},l_{\rm B},m_{\rm B},r,s,u}\partial_{\Theta_{\rm A}}\left[Z_{{\rm T}_{\rm A};l_{\rm A},m_{\rm A},L_{\rm I\!I}}\right]{\mathcal R}_{{\rm I\!I}; L_{\rm I\!I}}^{\rm Ad}\right\}}\, ,
\end{eqnarray}
\begin{eqnarray}
\lefteqn{I_{\rm A}\Omega_{\rm A}\frac{{\rm d}}{{\rm d}t}\cos\varepsilon_{\rm A}=M_{\rm C}\sum_{l_{\rm A},m_{\rm A},l_{\rm C},m_{\rm C},v,w,b}{\rm R}_{\rm e}\left\{-i\left[v+\left(m_{\rm A}+m_{\rm C}\right)\cos\varepsilon_{\rm A}\right]{\mathcal R}_{{\rm S}-{\rm S};l_{\rm A},m_{\rm A},l_{\rm C},m_{\rm C},v,w,b}\right.}\nonumber\\
& &-i\sum_{l_{\rm B},m_{\rm B},j,p,q}\left[\left(v-j\right)+\left(m_{\rm C}-m_{\rm B}\right)\cos\varepsilon_{\rm A}\right]{\mathcal R}_{{\rm I}; L_{\rm I}}-i\sum_{l^{'}_{\rm A},m^{'}_{\rm A},l_{\rm B},m_{\rm B},r,s,u}\left[\left(v-r\right)+\left(m_{\rm C}-\left(m^{'}_{\rm A}+m_{\rm B}\right)\right)\cos\varepsilon_{\rm A}\right]{\mathcal R}_{{\rm I\!I}; L_{\rm I\!I}}\nonumber\\
& &{\left.+\sum_{l_{\rm B},m_{\rm B},j,p,q}\left[\left(\partial_{\phi_{\rm A}}+\cos\varepsilon_{\rm A}\partial_{\Theta_{\rm A}}\right)Z_{{\rm T}_{\rm A};l_{\rm A},m_{\rm A},L_{\rm I}}\right]{\mathcal R}_{{\rm I}; L_{\rm I}}^{\rm Ad}+\sum_{l^{'}_{\rm A},m^{'}_{\rm A},l_{\rm B},m_{\rm B},r,s,u}\left[\left(\partial_{\phi_{\rm A}}+\cos\varepsilon_{\rm A}\partial_{\Theta_{\rm A}}\right)Z_{{\rm T}_{\rm A};l_{\rm A},m_{\rm A},L_{\rm I\!I}}\right]{\mathcal R}_{{\rm I\!I}; L_{\rm I\!I}}^{\rm Ad}\right\}}
\end{eqnarray}
and
\begin{eqnarray}
\lefteqn{\frac{1}{a_{\rm C}}\frac{{\rm d}a_{\rm C}}{{\rm d}t}=\frac{2}{n_{\rm C}}\frac{1}{a_{\rm C}^{2}}\sum_{l_{\rm A},m_{\rm A},l_{\rm C},m_{\rm C},v,w,b}{\rm R}_{\rm e}\left\{i\left(l_{\rm A}+l_{\rm C}-2w+b\right)\left[{\mathcal R}_{{\rm S}-{\rm S};l_{\rm A},m_{\rm A},l_{\rm C},m_{\rm C},v,w,b}+\sum_{l_{\rm B},m_{\rm B},j,p,q}{\mathcal R}_{{\rm I}; L_{\rm I}}+\sum_{l^{'}_{\rm A},m^{'}_{\rm A},l_{\rm B},m_{\rm B},r,s,u}{\mathcal R}_{{\rm I\!I}; L_{\rm I\!I}}\right]\right.}\nonumber\\
& &{\left.+\sum_{l_{\rm B},m_{\rm B},j,p,q}\partial_{M_{\rm C}}\left[Z_{{\rm T}_{\rm A};l_{\rm A},m_{\rm A},L_{\rm I}}\right]{\mathcal R}_{{\rm I}; L_{\rm I}}^{\rm Ad}+\sum_{l^{'}_{\rm A},m^{'}_{\rm A},l_{\rm B},m_{\rm B},r,s,u}\partial_{M_{\rm C}}\left[Z_{{\rm T}_{\rm A};l_{\rm A},m_{\rm A},L_{\rm I\!I}}\right]{\mathcal R}_{{\rm I\!I}; L_{\rm I\!I}}^{\rm Ad}\right\}}\, ,
\end{eqnarray}
\begin{eqnarray}
\lefteqn{\frac{1}{e_{\rm C}}\frac{{\rm d}e_{\rm C}}{{\rm d}t}=\frac{1}{n_{\rm C}}\frac{1-e_{\rm C}^{2}}{e_{\rm C}^{2}}\frac{1}{a_{\rm C}^{2}}}\nonumber\\
& & \times \sum_{l_{\rm A},m_{\rm A},l_{\rm C},m_{\rm C},v,w,b}{\rm R}_{\rm e}\left\{i\left[\left(l_{\rm A}+l_{\rm C}-2w\right)\left(1-\frac{1}{\sqrt{1-e_{\rm C}^{2}}}\right)+b\right]\left[{\mathcal R}_{{\rm S}-{\rm S};l_{\rm A},m_{\rm A},l_{\rm C},m_{\rm C},v,w,b}+\sum_{l_{\rm B},m_{\rm B},j,p,q}{\mathcal R}_{{\rm I}; L_{\rm I}}+\sum_{l^{'}_{\rm A},m^{'}_{\rm A},l_{\rm B},m_{\rm B},r,s,u}{\mathcal R}_{{\rm I\!I}; L_{\rm I\!I}}\right]\right.\nonumber\\
& &{\left.+\sum_{l_{\rm B},m_{\rm B},j,p,q}\left[\left(\partial_{M_{\rm C}}-\frac{1}{\sqrt{1-e_{\rm C}^{2}}}\partial_{\omega_{\rm C}}\right)Z_{{\rm T}_{\rm A};l_{\rm A},m_{\rm A},L_{\rm I}}\right]{\mathcal R}_{{\rm I}; L_{\rm I}}^{\rm Ad}+\sum_{l^{'}_{\rm A},m^{'}_{\rm A},l_{\rm B},m_{\rm B},r,s,u}\left[\left(\partial_{M_{\rm C}}-\frac{1}{\sqrt{1-e_{\rm C}^{2}}}\partial_{\omega_{\rm C}}\right)Z_{{\rm T}_{\rm A};l_{\rm A},m_{\rm A},L_{\rm I\!I}}\right]{\mathcal R}_{{\rm I\!I}; L_{\rm I\!I}}^{\rm Ad}\right\}}
\end{eqnarray}
\begin{eqnarray}
\lefteqn{\frac{{\rm d}}{{\rm d}t}\cos I_{\rm C}=\frac{1}{n_{\rm C}}\frac{1}{\sqrt{1-e_{\rm C}^{2}}}\frac{1}{a_{\rm C}^{2}}\sum_{l_{\rm A},m_{\rm A},l_{\rm C},m_{\rm C},v,w,b}{\rm R}_{\rm e}\left\{i\left[v-\left(l_{\rm A}+l_{\rm C}-2w\right)\cos I_{\rm C}\right]
\left[{\mathcal R}_{{\rm S}-{\rm S};l_{\rm A},m_{\rm A},l_{\rm C},m_{\rm C},v,w,b}+\sum_{l_{\rm B},m_{\rm B},j,p,q}{\mathcal R}_{{\rm I}; L_{\rm I}}+\sum_{l^{'}_{\rm A},m^{'}_{\rm A},l_{\rm B},m_{\rm B},r,s,u}{\mathcal R}_{{\rm I\!I}; L_{\rm I\!I}}\right]\right.}\nonumber\\
& &{\left.+\sum_{l_{\rm B},m_{\rm B},j,p,q}\left[\left(\partial_{{\Omega}_{\rm C}}-\cos I_{\rm C}\partial_{\omega_{\rm C}}\right)Z_{{\rm T}_{\rm A};l_{\rm A},m_{\rm A},L_{\rm I}}\right]{\mathcal R}_{{\rm I}; L_{\rm I}}^{\rm Ad}+\sum_{l^{'}_{\rm A},m^{'}_{\rm A},l_{\rm B},m_{\rm B},r,s,u}\left[\left(\partial_{\Omega_{\rm C}}-\cos I_{\rm C}\partial_{\omega_{\rm C}}\right)Z_{{\rm T}_{\rm A};l_{\rm A},m_{\rm A},L_{\rm I\!I}}\right]{\mathcal R}_{{\rm I\!I}; L_{\rm I\!I}}^{\rm Ad}\right\}}
\end{eqnarray}
where the explicit expression for ${\mathcal R}_{{\rm S}-{\rm S};l_{\rm A},m_{\rm A},l_{\rm C},m_{\rm C},v,w,b}$, ${\mathcal R}_{{\rm I}; L_{\rm I}}$ and ${\mathcal R}_{{\rm I\!I}; L_{\rm I\!I}}$ have been derived in Eqs. (\ref{RAC2}), (\ref{RAC5}) and (\ref{RAC6}).

\section{Scaling laws in the case of an extended axisymmetric deformed perturber}

\subsection{Comparison to the ponctual case}

Here, our goal is to quantify the term(s) of the disturbing function due to the non-ponctual behaviour of the perturber B and to compare it to the one in the ponctual mass case. 

To achieve this aim, some assumptions are assumed. First, we adopt the quadrupolar approximation for the response of A to the tidal excitation by B; thus, we put $l_{\rm A}=2$ so that ${\mathcal R}_{{\rm I\!I};L_{\rm I\!I}}=0$. Then, we consider the simplified situation where the body of which dynamics is studied is the tidal perturber: therefore B=C and Eq. (\ref{RAC5}) becomes:
\begin{eqnarray}
{\mathcal R}_{{\rm I};L_{\rm I}}&=&-\frac{G}{M_B}\frac{4\pi}{5}k_{2}^{\rm A}R_{\rm A}^5\left\vert Z_{{\rm T}_{\rm A};2,m_{\rm A},L_{\rm I}}\left(\nu,K;\Psi_{2+l_{\rm B},m_{\rm A}+m_{\rm B},j,p,q}\right) \right\vert \left[\gamma_{l_{\rm B},m_{\rm B}}^{2,m_{\rm A}}\right]^2\left\vert M_{l_{\rm B},m_{\rm B}}^{\rm B} \right\vert ^2\nonumber\\
& &\times\frac{1}{a_{\rm B}^{2\left(2+l_{\rm B}+1\right)}}\left[\kappa_{2+l_{\rm B},j}\right]^2\left[d_{j,m_{\rm A}+m_{\rm B}}^{2+l_{\rm B}}\left(\varepsilon_{\rm A}\right)\right]^{2}\left[F_{2+l_{\rm B},j,p}\left(I_{\rm B}\right)\right]^{2}\left[G_{2+l_{\rm B},p,q}\left(e_{\rm B}\right)\right]^{2}\nonumber\\
& &\times\exp\left[i\,\delta_{{\rm T}_{\rm A};2,m_{\rm A},L_{\rm I}}\left(\nu,K;\Psi_{2+l_{\rm B},m_{\rm A}+m_{\rm B},j,p,q}\right)\right]\,.
\end{eqnarray}
On the other hand, since we are interested in the amplitude of ${\mathcal R}_{{\rm I};L_{\rm I}}$, we focus on its norm ($\vert{\mathcal R}_{{\rm I};L_{\rm I}}\vert$). Finally, as we know that the dissipative part of the tide is very small compared to the adiabatic one (cf. Zahn 1966), we can assume that $\left\vert Z_{{\rm T}_{\rm A}}\right\vert\approx1$ in this first step.

\begin{figure}[h!]
\centering
\resizebox{12.75cm}{!}{\includegraphics{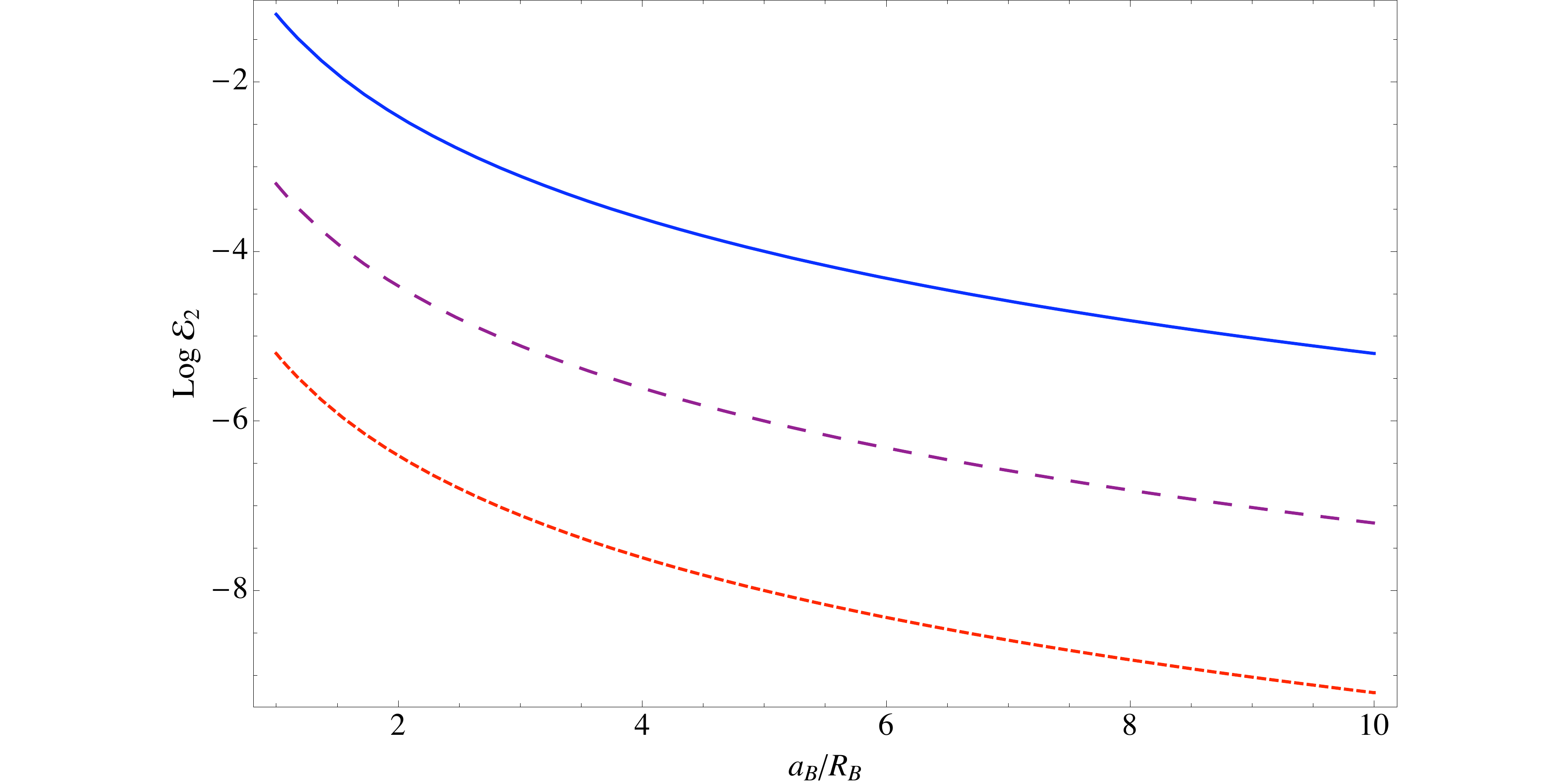}}
\caption{\bf $\rm Log\,{\mathcal E}_{2}$ in function of $a_{\rm B}/R_{\rm B}$ for $J_{2}=10^{-3}$ (red dashed line), $10^{-2}$ (purple long-dashed line), $10^{-1}$ (blue solid line). The non-ponctual terms have to be taken into account for strongly deformed perturbers ($J_{2}\ge10^{-2}$) in very close systems ($a_{\rm B}/R_{\rm B}\le5$) while they decrease rapidly otherwise.}
\label{f1}
\end{figure}

Let us first derive the term of $\vert{\mathcal R}_{{\rm I};L_{\rm I}}\vert$ due to the non-ponctual term of the gravific potential of B which has a non-zero average in time over an orbital period of B, $\left<V^{\rm B}_{\rm N-P}\right>_{T_{\rm B}}\left(\vec r\right)=1/T_{\rm B}\int_{0}^{T_{\rm B}}V^{\rm B}_{\rm N-P}\left(t,\vec r\right){\rm d}t$ that corresponds to the axisymmetric rotational and permanent tidal deformations (see Zahn 1977) (the same procedure can of course be applied to the non-stationnary and non-axisymmetric deformations, but we choose here to focus only on $\left<V^{\rm B}_{\rm N-P}\right>_{T_{\rm B}}$ to illustrate our purpose). Then, as the considered deformations of B are axisymmetric, we can expand them using the usual gravitational moments of B ($J_{l_B}$) as
\begin{equation}
V^{\rm B}\left(\vec r\right)=\frac{GM_{\rm B}}{r}+\left<V^{\rm B}_{\rm N-P}\right>_{T_{\rm B}}\,\,\,\hbox{where}\,\,\,\left<V^{\rm B}_{\rm N-P}\right>_{T_{\rm B}}=G\sum_{l_{\rm B}>0}\left(M_{l_{\rm B},0}^{\rm S_{\rm B}}+M_{l_{\rm B},0}^{\rm T_{\rm B}}\right)\frac{Y_{l_{\rm B},0}\left(\theta,\varphi\right)}{r^{l_{\rm B}+1}}\,\,\,\hbox{with}\,\,\,
M_{l_{\rm B},0}^{\rm S_{\rm B}}+M_{l_{\rm B},0}^{\rm T_{\rm B}}=-\frac{J_{l_{\rm B}} M_{\rm B} R_{\rm B}^{l_{\rm B}}}{{\mathcal N}_{l_{\rm B}}^{0}}.
\end{equation}
Then, we obtain
\begin{equation}
\left\vert{\mathcal R}_{{\rm I};L_{\rm I}}^{J_{l_{\rm B}}}\left(a_{\rm B},e_{\rm B},I_{\rm B},\varepsilon_{\rm A}\right)\right\vert=\frac{G}{M_B}\frac{4\pi}{5}k_{2}^{\rm A}R_{\rm A}^5 \left[\gamma_{l_{\rm B},0}^{2,m_{\rm A}}\right]^2\left\vert M_{l_{\rm B},0}^{{\rm S}_{\rm B}} + M_{l_{\rm B},0}^{{\rm T}_{\rm B}} \right\vert ^2\frac{1}{a_{\rm B}^{2\left(2+l_{\rm B}+1\right)}}\left[\kappa_{2+l_{\rm B},j}\right]^2\left[d_{j,m_{\rm A}}^{2+l_{\rm B}}\left(\varepsilon_{\rm A}\right)\right]^{2}\left[F_{2+l_{\rm B},j,p}\left(I_{\rm B}\right)\right]^{2}\left[G_{2+l_{\rm B},p,q}\left(e_{\rm B}\right)\right]^{2}.
\label{App1}
\end{equation}
On the other hand, the term of $\vert{\mathcal R}_{{\rm I};L_{\rm I}}\vert$ associated to $M_{\rm B}$, namely the disturbing function in the case where B is assumed to be a ponctual mass, is given by:
\begin{eqnarray}
\lefteqn{\left\vert{\mathcal R}_{{\rm I};L_{\rm I}}^{M_{\rm B}}\left(a_{\rm B},e_{\rm B},I_{\rm B},\varepsilon_{\rm A}\right)\right\vert=\frac{G}{M_B}\frac{4\pi}{5}k_{2}^{\rm A}R_{\rm A}^5 \left[\gamma_{0,0}^{2,m_{\rm A}}\right]^2\left\vert M_{0,0}^{{\rm S}_{\rm B}} \right\vert ^2\frac{1}{a_{\rm B}^{6}}\left[\kappa_{2,j}\right]^2\left[d_{j,m_{\rm A}}^{2}\left(\varepsilon_{\rm A}\right)\right]^{2}\left[F_{2,j,p}\left(I_{\rm B}\right)\right]^{2}\left[G_{2,p,q}\left(e_{\rm B}\right)\right]^{2}}\nonumber\\
&=&\frac{G}{M_B}\frac{4\pi}{5}k_{2}^{\rm A}R_{\rm A}^5 M_{\rm B} \frac{1}{a_{\rm B}^{6}}\left[\kappa_{2,j}\right]^2\left[d_{j,m_{\rm A}}^{2}\left(\varepsilon_{\rm A}\right)\right]^{2}\left[F_{2,j,p}\left(I_{\rm B}\right)\right]^{2}\left[G_{2,p,q}\left(e_{\rm B}\right)\right]^{2}\,.
\label{App2}
\end{eqnarray}
For this first evaluation of the ratio $\left\vert{\mathcal R}_{{\rm I};L_{\rm I}}^{J_{l_{\rm B}}}\right\vert / \left\vert{\mathcal R}_{{\rm I};L_{\rm I}}^{M_{\rm B}}\right\vert$, we focus on the configuration of minimum energy. In this state, the spins of A and B are aligned with the orbital one so that $\varepsilon_{\rm A}=I_{\rm B}=0$ (that leads to $j=m_{\rm A}$ and $p=\left(2-m_{\rm A}+l_{\rm B}\right)/2$) and the orbit is circular ($e_{\rm B}=0$). Then, we consider:
\begin{equation}
{\mathcal E}_{m_{\rm A},l_{\rm B}}=\frac{\left\vert{\mathcal R}_{{\rm I};L_{\rm I}}^{J_{l_{\rm B}}}\left(a_{\rm B},0,0,0\right)\right\vert}{\left\vert{\mathcal R}_{{\rm I};L_{\rm I}}^{M_{\rm B}}\left(a_{\rm B},0,0,0\right)\right\vert}.
\end{equation}
Using eqs. (\ref{App1}-\ref{App2}), we get its expression in function of $J_{l_{\rm B}}$ and of $\left(R_{\rm B}/a_{\rm B}\right)$: 
\begin{equation}
{\mathcal E}_{m_{\rm A},l_{\rm B}}=\frac{1}{4\pi}\left[\frac{1}{{\mathcal N}_{l_{\rm B}}^{0}}\frac{\gamma_{l_{\rm B},0}^{2,m_{\rm A}}}{\gamma_{0,0}^{2,m_{\rm A}}}\frac{\kappa_{2+l_{\rm B},m_{\rm A}}}{\kappa_{2,m_{\rm A}}}\frac{F_{2+l_{\rm B},2,\frac{l_{\rm B}}{2}}\left(0\right)}{F_{2,2,0}\left(0\right)}\right]^2J_{l_{\rm B}}^{2}\left(\frac{R_{\rm B}}{a_{\rm B}}\right)^{2 l_{\rm B}}\,.
\end{equation}
As it has been emphasized by Zahn (1966-1977), the main mode of the dissipative tide ruling the secular evolution of the system is $m_{\rm A}=2$. We thus define ${\mathcal E}_{l_{\rm B}}$ such that
\begin{equation}
{\mathcal E}_{l_{\rm B}}={\mathcal E}_{2,l_{\rm B}}=\left[\frac{1}{3}F_{2+l_{B},2,\frac{l_{B}}{2}}\left(0\right)\right]^{2}J_{l_{\rm B}}^{2}\left(\frac{R_{\rm B}}{a_{\rm B}}\right)^{2 l_{\rm B}}\,,
\end{equation}
which can be recast into
\begin{equation}
\log\left({\mathcal E}_{l_B}\right)=2\left[\log\left[\frac{1}{3}F_{2+l_{B},2,\frac{l_{B}}{2}}\left(0\right)\right]+\log J_{l_B}-l_{B}\log\left(\frac{a_B}{R_B}\right)\right]\,.
\end{equation}
Finally, keeping only into account the quadrupolar deformation of B ($J_{2}$), we get:
\begin{equation}
\log\left({\mathcal E}_{2}\right)=2\left[\log\left(\frac{5}{2}\right)+\log J_{2}-2\log\left(\frac{a_B}{R_B}\right)\right]\,.
\label{e2}
\end{equation}
This gives us the order of magnitude of the terms due to the non-ponctual behaviour of B compared to the one obtained in the ponctual mass approximation. It is directly proportional to the squarred $J_{2}$, thus increasing with $\varepsilon_{\Omega}^2$ (where $\varepsilon_{\Omega}={\Omega_{\rm B}^2}/{\Omega_{\rm c}^{2}}$ with $\Omega_{\rm c}=\sqrt{\frac{G M_{\rm B}}{R_{\rm B}^{3}}}$) in the case of the rotation-induced deformation and with $\varepsilon_{\rm T}^{2}$ (where $\varepsilon_{\rm T}=q\left({R_{\rm B}}/{a_{\rm B}}\right)^3$ where $q=M_{\rm A}/M_{\rm B}$) in the tidal one, while it increases as $\left(R_{\rm B}/a_{\rm B}\right)^{4}$. Therefore, as it is shown in Fig. (\ref{f1}), the non-ponctual terms have to be taken into account for strongly deformed perturbers ($J_{2}\ge10^{-2}$) in very close systems ($a_{\rm B}/R_{\rm B}\le5$) while they decrease rapidly otherwise. {\bf This corresponds for example to the case of internal natural satellites of rapidly rotating giant planets as Jupiter and Saturn ($J_2\approx1.4697\cdot10^{-2}$ for Jupiter and $J_2\approx1.6332\cdot10^{-2}$ for Saturn; see Guillot 1999 and references therein). In the case of close Hot-Jupiters which are already synchronized (because of the tidal dissipation the rotation period is close to the orbital one), the rotation period is larger than 2 days (50 hours) that is roughly 5 times slower than Jupiter's rotation (10 hours). In this case, the flattening of Hot-Jupiter is less important and their $J_2$ should be of the same order of the Earth's value ({\it i.e.} $J_2$ runs from $10^{-4}$ to $10^{-3}$). Then, the relative effect of the non-ponctual terms is less important. The situation may be different in the earliest evolutionary stages of those systems and this have to be studied in forthcoming works.}

\subsection{Application to rapidly rotating binary stars}
Here, we apply the previous procedure to the case of rapidly rotating binary stars where 
$\varepsilon_{\Omega}\!>\!\!>\!\varepsilon_{\rm T}$. 

First, we roughly scale $J_{2}$ of B to $J_{\odot}$ as (cf. Roxburgh 2001):
\begin{equation}
J_{2}
\propto{\mathcal F}\left(\varepsilon_{\rm \Omega}\right)
\approx J_{2}^{\,\odot}\left(\frac{R_{\rm B}}{R_{\odot}}\right)^{3}\left(\frac{\Omega_{\rm B}}{\Omega_{\odot}}\right)^{2}\left(\frac{M_{\rm B}}{M_{\odot}}\right)^{-1}\,,
\end{equation}
the deformation being thus mainly due to rotation.
Then, Eq. (\ref{e2}) becomes
\begin{equation}
\log\left({\mathcal E}_{2}\right)\approx2\left[\log\left(\frac{5}{2}\right)+\log J_{2}^{\,\odot}+3\log\left(\frac{R_B}{R_{\odot}}\right)-\log\left(\frac{M_B}{M_{\odot}}\right)+2\log\left(\frac{\Omega_{B}}{\Omega_{\odot}}\right)-2\log\left(\frac{a_B}{R_B}\right)\right]\,.
\end{equation}
Then, we introduce the stellar homology relations for main-sequence stars (cf. Kippenhahn \& Weigert 1990):
\begin{equation}
\frac{R}{R_{\odot}}=\left(\frac{M}{M_{\odot}}\right)^{z_{1}}\,\,\,\hbox{where}\,\,\,z_{1}=\frac{\alpha+\lambda-2}{\alpha+3\lambda}\,\,\,\hbox{with}\,\,\,\lambda=\left(\frac{\partial\ln\epsilon}{\partial\ln\rho}\right)_{T}\,\,\,\hbox{and}\,\,\,\alpha=\left(\frac{\partial\ln\epsilon}{\partial\ln T}\right)_{\rho},
\end{equation}
$\epsilon$ being the nuclear energy production rate per unit mass inside the star. We finally obtain:
\begin{equation}
\log\left({\mathcal E}_{2}\right)\approx2\left[\log\left(\frac{5}{2}\right)+\log J_{2}^{\,\odot}+\left(3z_{1}-1\right)\log\left(\frac{M_B}{M_{\odot}}\right)+2\log\left(\frac{\Omega_{B}}{\Omega_{\odot}}\right)-2\log\left(\frac{a_B}{R_B}\right)\right].
\end{equation}
For Main-Sequence stars burning their hydrogen $\lambda\approx1$ while considering the pp chain leads to $z_{1}\approx0.465$ ($M_{\rm B}/M_{\odot}\le 1.5$) and the CNO cycle to $z_{1}\approx0.795$ ($M_{\rm B}/M_{\odot}\ge 1.5$) (see Kippenhahn \& Weigert 1990 for a detailed discussion). 

\begin{figure}[h!]
\centering
\resizebox{12.75cm}{!}{\includegraphics{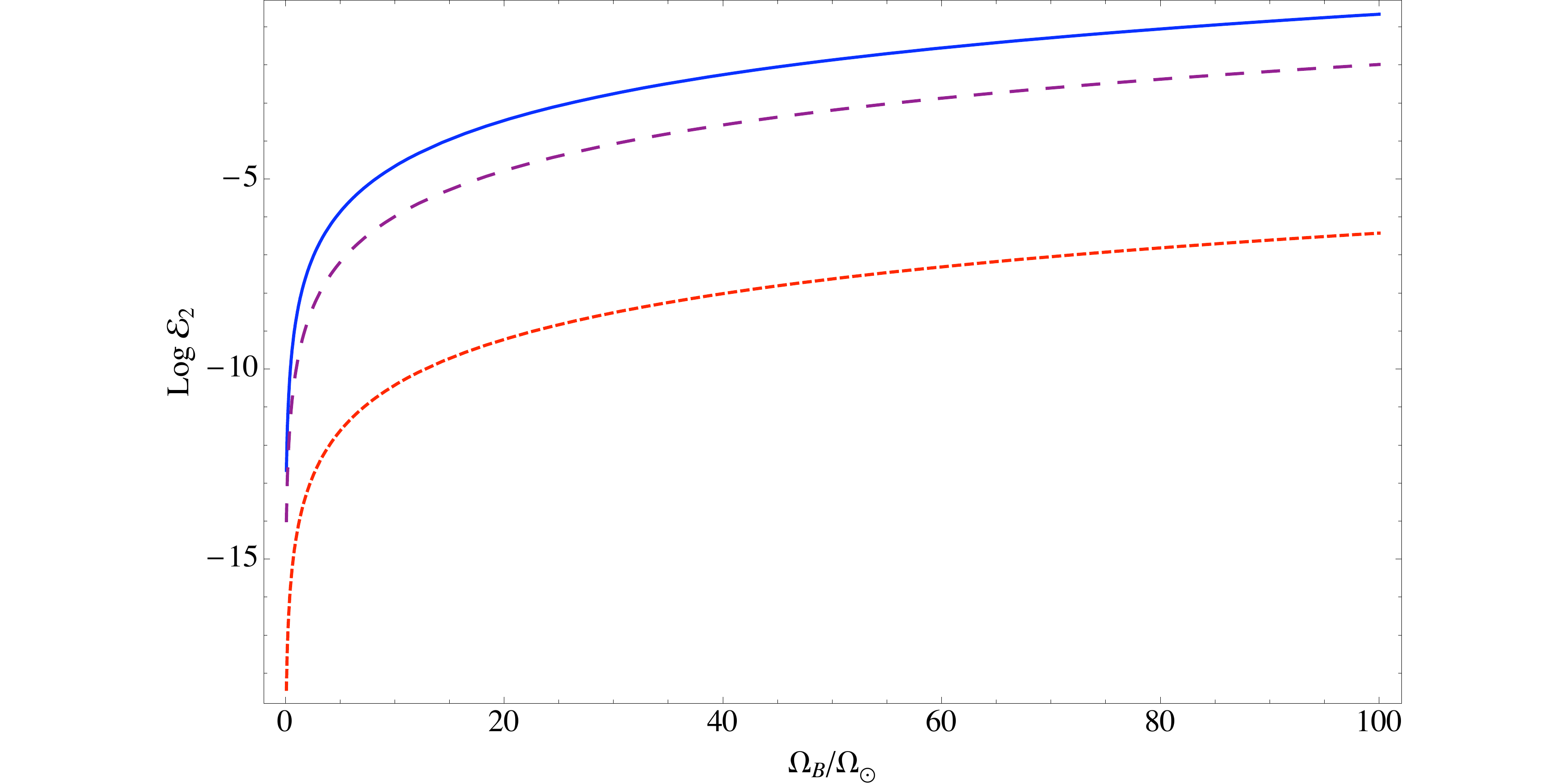}}
\caption{{\bf$\rm Log\,{\mathcal E}_{2}$ for 1$M_{\odot}$ (red dashed line), 40$M_{\odot}$ (purple long-dashed line) and 120$M_{\odot}$ (solid blue line) Main-Sequence stars in function of %$M_{\rm B}/M_{\odot}$ and of 
$\Omega_{\rm B}/\Omega_{\odot}$ at $a_{\rm B}=3\,R_{\rm B}$ taking $J_{2}^{\,\odot}\approx2.2\, 10^{-7}$. Non-ponctual terms are not always a perturbation compared to the ponctual mass one (${\mathcal E}_{2}\ge10^{-2}$) for $M_{\rm B}\ge40\,M_{\odot}$ and $\Omega_{\rm B}\ge45\,\Omega_{\odot}$ that corresponds to rapidly rotating close binary massive stars.}}
\label{f2}
\end{figure}

In Fig. (\ref{f2}), we plot ${\mathcal E}_{2}$ in function of $M_{\rm B}/M_{\odot}$ and $\Omega_{\rm B}/\Omega_{\odot}$ for $a_{\rm B}=3\,R_{\rm B}$ taking $J_{2}^{\,\odot}\approx2.2\, 10^{-7}$. It is shown that the non-ponctual terms are not always a perturbation compared to the ponctual mass one (${\mathcal E}_{2}\ge10^{-2}$) for $M_{\rm B}\ge40\,M_{\odot}$ and $\Omega_{\rm B}\ge45\,\Omega_{\odot}$ that corresponds to rapidly rotating close binary massive stars. This has to be taken into account in future studies dedicated to the evolution of such stars.

Since, we have now identified the regime where the non-ponctual terms have to be taken into account, those have to be examined by integrating the complete set of dynamical equations. This will be presented in a forthcoming paper.

\section{Conclusion}

\par This work represents another step towards the modelling of the tidal dynamical evolution of planetary systems and of multiple stars by taking into account the extended character of the bodies. We have used the STF tensors which allow us to treat in an analytical and compact way the complex couplings between the multipole behaviour of gravitational fields of two extended bodies. First, the main properties of STF tensors have been defined and derived. Next, after deriving the classical multipole expansion of the self gravitational field of an extended body A, we have derived  the tidal potential related to the interaction between two extended bodies A and B, which is a generalization of the previous results where the perturber is treated through the ponctual mass approximation. Using those results, the external potential of a tidally-perturbed body has been provided and used to obtain the associated mutual interaction potential with a body C for which the dynamics is studied and which could be different from the previous tidal perturber B, as well as the corresponding disturbing function. The disturbing function being expanded into Fourier's series, the dynamical equations are derived. Those allows us to study the dynamical evolution of the angular momentum and the obliquity of the body A simultaneously with the Keplerian oribtal elements of C relative to the center of mass of A, namely the semi major axis, $a_{\rm C}$, the eccentricity, $e_{\rm C}$, and the orbital inclination, $I_{\rm C}$. {The equations have been derived in a general way without any linearization of function of any orbital elements as it has been done in Hut (1981-1982) and in Correia \& Laskar (2003c).} This allows us to study the strongly non-linear problem of inclined, eccentric orbits. Therefore, this formalism can be useful for elliptic systems.\\

\par The major interest of this modelling concerns the dissipation of the tidal mechanical energy. Here, the response of each extended body is parametrized through the Love number $k_{l}$ for the adiabatic one, and an impedance $Z_{\rm T}$ and its associated delay $\delta_{\rm T}$ which describes the damping and the phase lag due to the (viscous and thermal) dissipation acting on the bulk induced by the perturber. Self-consistent modellings have to be developed for elasto-viscous bodies (Rogister and Rochester, 2004) as well as for fluid bodies (cf. Zahn, 1977) as well as equilibrium and stability conditions. This will be done for extended fluid bodies in a forthcoming sery of papers.\\

\par Moreover, due to the general character of this work, various applications are possible. The first one deals with the dynamics of planets which are very close to their star as observed in several extrasolar planetary systems (cf. Laskar \& Correia, 2004; Levrard et al., 2007; Fabrycky et al., 2007; {\bf Correia et al. 2008}). Next, in our own Solar System, dynamical systems, such as giant planets and their internal satellites, can be studied by taking into account their spatial extension and their multipolar behaviour due for example to their rapid rotation. Finally, this work can be applied to the dynamical evolution of close binary stars where the tidal interaction and its dissipation dominate the behaviour of the system until it has reached its lower energy state, where the spins are all aligned, the orbits are circular, and the components are synchronized with the orbital motion. For these coming applications, the potential impact of the spatial extension of bodies on dynamics has to be carefully evaluated and understood.

\begin{acknowledgement}
This is a real pleasure to acknowledge Pr. J.-P. Zahn, Pr. T. Damour and Dr. J. Vaubaillon for helpful comments. We are also grateful to the anonymous referee for his suggestions that allow us to improve the original manuscript.
\end{acknowledgement}


\begin{thebibliography}{}

\bibitem{} Abramowitz, M., Stegun, I. 1972, Handbook of mathematical functions, Dover, New York

\bibitem{} Alexander, M. E. 1973, Astrophysics and Space Science, 23, 459

\bibitem{} Andoyer,  H. 1926, M\'ecanique C\'eleste, Gauthier-Villars, Paris

\bibitem{} Blanchet, L., Damour, T. 1986, Phil. Trans. R. Soc. Lon., A 320, 379

\bibitem{} Bord\'e, P., Rouan, D., L\'eger, A. 2003, 405, 1137

\bibitem{} Borderies, N. 1978, Celestial Mechanics, 18, 295

\bibitem{} Borderies, N. 1980, A\&A, 82, 129

\bibitem{} Borderies, N., Yoder, C. F. 1990, A\&A, 233, 235

\bibitem{} Borucki, W. J. et al. 2007, Transiting Extrasolar Planets, Proceedings of the conference held 25-28 September 2006 at the Max Planck Institute for Astronomy in Heidelberg, Germany; Eds: C. Afonso, D. Weldrake and Th. Henning; San Francisco Astronomical Society of the Pacific; ASP Conference Series, 366, 309

\bibitem{} Brouwer, D., Clemence, G. M. 1961, Methods of celestial mechanics, New York: Academic Press

\bibitem{} Correia, A.C.M. 2001, Ph. D. Thesis, Universit\'e Paris VII

\bibitem{} Correia, A.C.M., Laskar, J. 2001, Nature (Issue 6839), 411, 767

\bibitem{} Correia, A.C.M., Laskar, J., N\'eron de Surgy, O. 2003a, Icarus, 163, 1

\bibitem{} Correia, A.C.M., Laskar, J. 2003b, Icarus, 163, 24

\bibitem{} Correia, A.C.M., Laskar, J. 2003c, Journal of Geophysical Research, 108 (Issue. E11), 5123

\bibitem{} {\bf Correia, A.C.M., Levrard, B., Laskar, J. 2008, A\&A, 488, L63} 

\bibitem{} Courant, R. \& Hilbert, D. 1953, Methods of Mathematical Physics, Interscience, New York

\bibitem{} Damour, T., Soffel, M. H., Xu, C. 1992, Phys. Rev. D, 45, 1017

\bibitem{} Fabrycky, D. C., Johnson, E. T., Goodman, J. 2007, ApJ, 665, 754

\bibitem{} Gelfand, I. M., Minlos, R. A., Shapiro, Z. Ya. 1963, Representation of the Rotation of the Lorentz groups, Pergamon, Oxford

\bibitem{} {\bf Guillot, T. 1999, \planss, 47, 1183}

\bibitem{} Hartmann, T., Soffel, M. H., Kioustelidis, T. 1994, Celestial Mechanics and Dynamical Astronomy, 60, 139

\bibitem{} Hut, P. 1981, A\&A, 99, 126

\bibitem{} Hut, P. 1982, A\&A, 110, 37

\bibitem{} Ilk, K. H. 1983, Ph. D. Thesis, Technischen Universitaet, Munich Bayerische Akademie der Wissenschaften

\bibitem{} Kaula, W. M. 1962, The Astronomical Journal, 67, 300

\bibitem{} Kippenhahn, R., Weigert, W. 1990, Stellar Structure and Evolution, Springer

\bibitem{} Lambeck, K. 1980, The Earth's Variable Rotation, Cambridge University Press

\bibitem{} Levrard, B., et al. 2007, A\&A, 462, L5

\bibitem{} Laskar, J., Correia, A.C.M. 2004, Extrasolar Planets: Today and Tomorrow, Eds: Jean-Philippe Beaulieu, Alain Lecavelier des Etangs and Caroline Terquem, ASP Conference Proceedings, 321, 401

\bibitem{} Laskar, J. 2005, Celestial Mechanics and Dynamical Astronomy, 91, 351

\bibitem{} Maciejewski, A. J. 1995, Celestial Mechanics and Dynamical Astronomy, 63, 1

\bibitem{} Mayor, M., Pont, F., Vidal-Madjar, A. 2005, Progress of Theoretical Physics Supplement, 158, 43

\bibitem{} Melchior, P. J. 1971, Physique et dynamique plan\'etaire, Vander, Bruxelles

\bibitem{} N\'eron de Surgy, O. 1996, Ph. D. Thesis, Observatoire de Paris

\bibitem{} N\'eron de Surgy, O., Laskar, J. 1997, A\&A, 318, 975

\bibitem{} Rogister, Y., Rochester, M. G. 2004, Geophysical Journal International, 159, 874

\bibitem{} Roxburgh, I. W. 2001, A\&A, 377, 688

\bibitem{} Thorne, K. 1980, Rev. Mod. Phys., 52, 299

\bibitem{} Tisserand, F. F. 1889, Trait\'e de M\'ecanique C\'eleste Tome I, Paris, Gauthier Villars

\bibitem{} Tisserand, F. F. 1891, Trait\'e de M\'ecanique C\'eleste Tome II, Paris, Gauthier Villars

\bibitem{} Yoder, C. F. 1995, Icarus, 117, 250

\bibitem{} Zahn, J.-P. 1966, Annales d'Astrophysique, 29, 313

\bibitem{} Zahn, J.-P. 1966, Annales d'Astrophysique, 29, 489

\bibitem{} Zahn, J.-P. 1977, A\&A, 57, 383

\end{thebibliography}
\end{document}